\documentclass[aip,reprint]{revtex4-1}
\usepackage{gensymb}
\usepackage{graphicx}
\usepackage{caption}
\usepackage{subcaption}
\usepackage{amsmath}
\usepackage{soul}
\usepackage{color}
\usepackage{upgreek}

\draft 

\begin{document}

\title{Tuning of perpendicular magnetic anisotropy in Bi-substituted yttrium iron garnet films by He$^{+}$ ion irradiation} 

\author{Sreeveni Das}
\author{Rhodri Mansell}
\email[Corresponding author: ]
{rhodri.mansell@aalto.fi}
\author{Lukáš Flajšman}
\affiliation{NanoSpin, Department of Applied Physics, Aalto University School of Science, PO Box 15100, FI-00076 Aalto, Finland}

\author{Lide Yao}
\affiliation{OtaNano-Nanomicroscopy Center, Aalto University, PO Box 15100, FI-00076 Aalto, Finland\looseness=-1}

\author{Johannes W. van der Jagt}
\author{Song Chen}
\affiliation{Spin-Ion Technologies, 10 Boulevard Thomas Gobert, 91120 Palaiseau, France}
\affiliation{
Université Paris-Saclay, 3 rue Joliot Curie, Gif-sur-Yvette 91190, France}
\author{Dafiné Ravelosona}
\affiliation{Spin-Ion Technologies, 10 Boulevard Thomas Gobert, 91120 Palaiseau, France}
\affiliation{Centre de Nanosciences et de Nanotechnologies, CNRS, Université Paris-Saclay, 10 Boulevard Thomas Gobert, Palaiseau 91120, France}

\author{Liza Herrera Diez}
\affiliation{Centre de Nanosciences et de Nanotechnologies, CNRS, Université Paris-Saclay, 10 Boulevard Thomas Gobert, Palaiseau 91120, France}

\author{Sebastiaan van Dijken}
\email[Corresponding author: ]{sebastiaan.van.dijken@aalto.fi}
\affiliation{NanoSpin, Department of Applied Physics, Aalto University School of Science, PO Box 15100, FI-00076 Aalto, Finland}

\begin{abstract}
We report the continuous tuning of magnetic anisotropy in perpendicularly magnetized bismuth-substituted yttrium iron garnet (Bi-YIG) films via He$^+$ ion irradiation. Our findings indicate that the magnetization direction of epitaxial Bi-YIG films on sGGG substrates transitions from out-of-plane in the as-grown state to in-plane after He$^+$ ion irradiation at a fluence exceeding $2\times10^{14}$ ions/cm$^2$. The reorientation is attributed to the relaxation of tensile film strain, which reduces the perpendicular magnetic anisotropy without affecting the saturation magnetization. The Gilbert damping parameter and the inhomogeneous broadening of the ferromagnetic resonance linewidth show only minimal increases with ion irradiation. Additionally, at a fluence of $5\times10^{13}$ ions/cm$^2$, we observe the formation of magnetic bubble domains in the Bi-YIG films. Micromagnetic simulations estimate a Dzyaloshinskii-Moriya interaction of 0.006 mJ/m\(^2\), which is insufficient for stabilizing N\'{e}el-type skyrmions. Finally, we demonstrate that the effects of He$^+$ ion irradiation can be largely reversed through thermal annealing in an oxygen atmosphere. 
\end{abstract}


\maketitle 

Control over perpendicular magnetic anisotropy (PMA) and the Dzyaloshinskii–Moriya interaction (DMI) in magnetic films is crucial for spintronic devices that utilize chiral magnetic textures such as N\'{e}el domain walls and skyrmions. In insulating magnetic garnets, such as yttrium iron garnet (YIG)\cite{Ding2020, Fu2017, Chen2022}, substituted-YIG\cite{Wang2017a, Lage2017, Soumah2018, Liu2019, Lin2020, Kuila2022, Boettcher2022, Jia2023, Das2023}, and rare-earth iron garnets\cite{Wu2018, Rosenberg2018, Ortiz2018, Bauer2019}, PMA has been induced through epitaxial film growth onto lattice-mismatched garnet substrates. Robust PMA in iron garnets is of particular interest because it facilitates fast magnetization dynamics via spin-orbit torque (SOT) switching \cite{Avci2016}, offers high domain wall velocities in magnetic racetrack devices\cite{Velez2019, Avci2019}, and allows the isotropic excitation of magnetostatic forward volume mode (MSFVM) spin waves in magnon-based computing devices\cite{Klingler2015}. 

Chiral spin textures\cite{Shao2019} and DMI have been reported in YIG and various rare-earth iron garnets, such as thulium iron garnet (TmIG)\cite{Caretta2020, Wang2020b, Schlitz2021}. Moreover, bubble-like topological textures have been imaged in TmIG films, with or without a heavy-metal overlayer\cite{Ding2019, Buttner2020}, and current-driven dynamics of skyrmion bubbles has been observed in YIG/TmIG bilayers\cite{Velez2022}. While these findings might suggest that rare-earth elements are essential for establishing DMI in garnet films\cite{Caretta2020}, this assertion is contradicted by the observation of relatively large DMI in YIG/GGG\cite{Wang2020b}. Therefore, the origin of DMI in garnet films remains complex and not fully understood. 
 
YIG and substituted-YIG films exhibit lower magnetic damping compared to rare-earth iron garnets, making the tailoring of PMA and DMI in these materials particularly relevant. Bismuth-substituted YIG (Bi-YIG) films have been found to combine robust PMA and low magnetic damping\cite{Soumah2018, Liu2019, Lin2020, Jia2023, Das2023}. Previously, however, zero DMI has been measured in single Bi-YIG films\cite{Caretta2020}, and DMI has only been reported for Bi-YIG/TmIG bilayers on NGG substrates\cite{Fakhrul2024}. 

Irradiation of magnetic films and bilayers with light ions has been employed to modify interface effects such as PMA\cite{Sakamaki2012,Balan2023}, DMI\cite{Diez2019,Juge2021,Gieniusz2021,Nembach2022}, exchange bias\cite{Mewes2000}, and interlayer exchange coupling\cite{Teixeira2020}. While the effects of ion irradiation on micrometer-thick garnets have been extensively studied\cite{Suzuki1986,Guzman1983}, there are only a few reports on the impact of ion irradiation on the magnetic properties of nanometer-thick garnet films\cite{Ruane2017,Kiechle2023}. For spintronic applications, precise control over PMA and DMI in thin films is crucial. 

\begin{figure*}[htp]
\centering
    \includegraphics[width=1.0\linewidth]{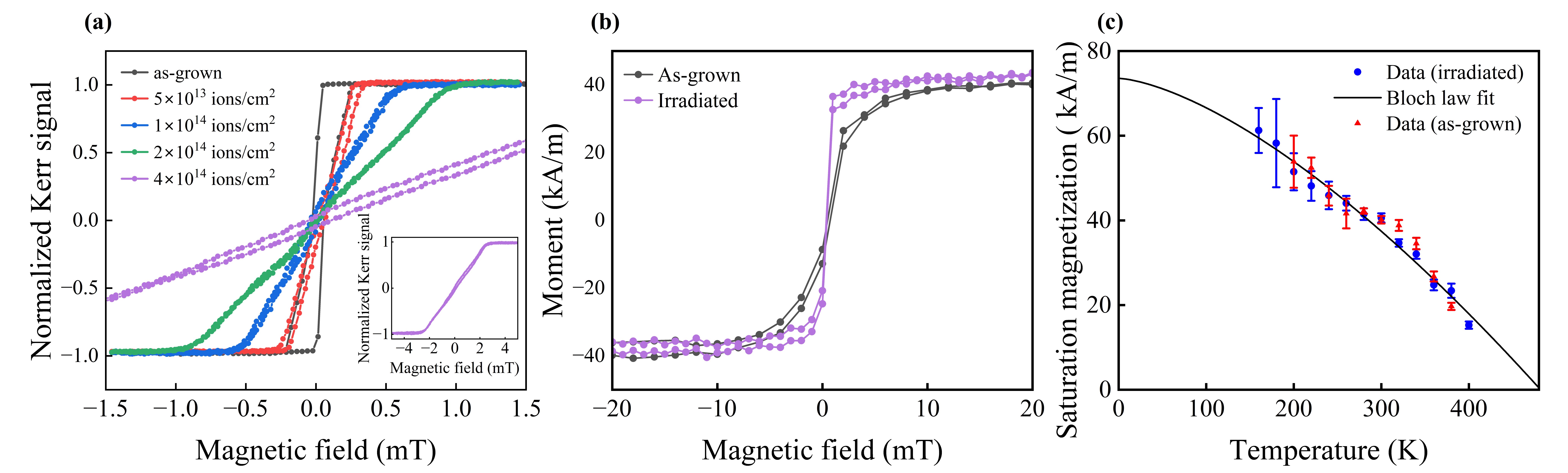}
    \caption{(a) Polar MOKE hysteresis loops of the Bi-YIG film measured under perpendicular applied magnetic field in the as-grown state and after He\(^+\) ion irradiation with different fluences. The inset shows the zoomed-out hysteresis loop for the film with the highest irradiation. (b) VSM hysteresis curves recorded with an in-plane magnetic field at 300 K for an as-grown Bi-YIG film and the Bi-YIG film after ion irradiation with a fluence of 4\(\times\)10\(^{14}\) ions/cm\(^2\). (c) Temperature dependence of \(M_\mathrm{s}\) for the same films as in (b). The black line represents a Bloch law fit to the data recorded on the film after ion irradiation.}
    \label{moke2}
\end{figure*}

In this study, we utilize He$^+$ ion irradiation as a post-deposition process to tailor the properties of perpendicularly magnetized Bi-YIG films. Light He$^+$ ions are chosen, rather than heavier ions such as Ga$^+$, so that the ion implantation occurs through the film thickness and is not dominated by surface effects associated with the shorter stopping distance of heavier ions. We demonstrate that the magnetization of the films is reoriented from out-of-plane to in-plane by He$^+$ ion irradiation, while the Gilbert damping parameter and the inhomogeneous broadening of the ferromagnetic resonance (FMR) linewidth are minimally affected. Using magneto-optical Kerr effect (MOKE) microscopy, we observe the formation of magnetic bubble domains in the Bi-YIG films after ion irradiation with a small fluence. To evaluate the chirality of these bubbles, we assess the strength of DMI before and after multiple ion-irradiation steps, finding that the DMI remains small and constant. The reorientation of magnetization and the stabilization of bubble domains in the Bi-YIG films are driven by a reduction in PMA due to ion-induced strain relaxation. Thermal annealing of ion-irradiated Bi-YIG films in an oxygen atmosphere restores most of the PMA. This study highlights the potential of He$^+$ irradiation for controlling of the properties of ultrathin Bi-YIG films and shows that such films can indeed host a substantial DMI. The results demonstrate a promising approach for combining chiral magnetic domains with low-damping perpendicularly magnetized materials.

\begin{figure*}[btp]
\centering
    \includegraphics[width=1.0\linewidth]{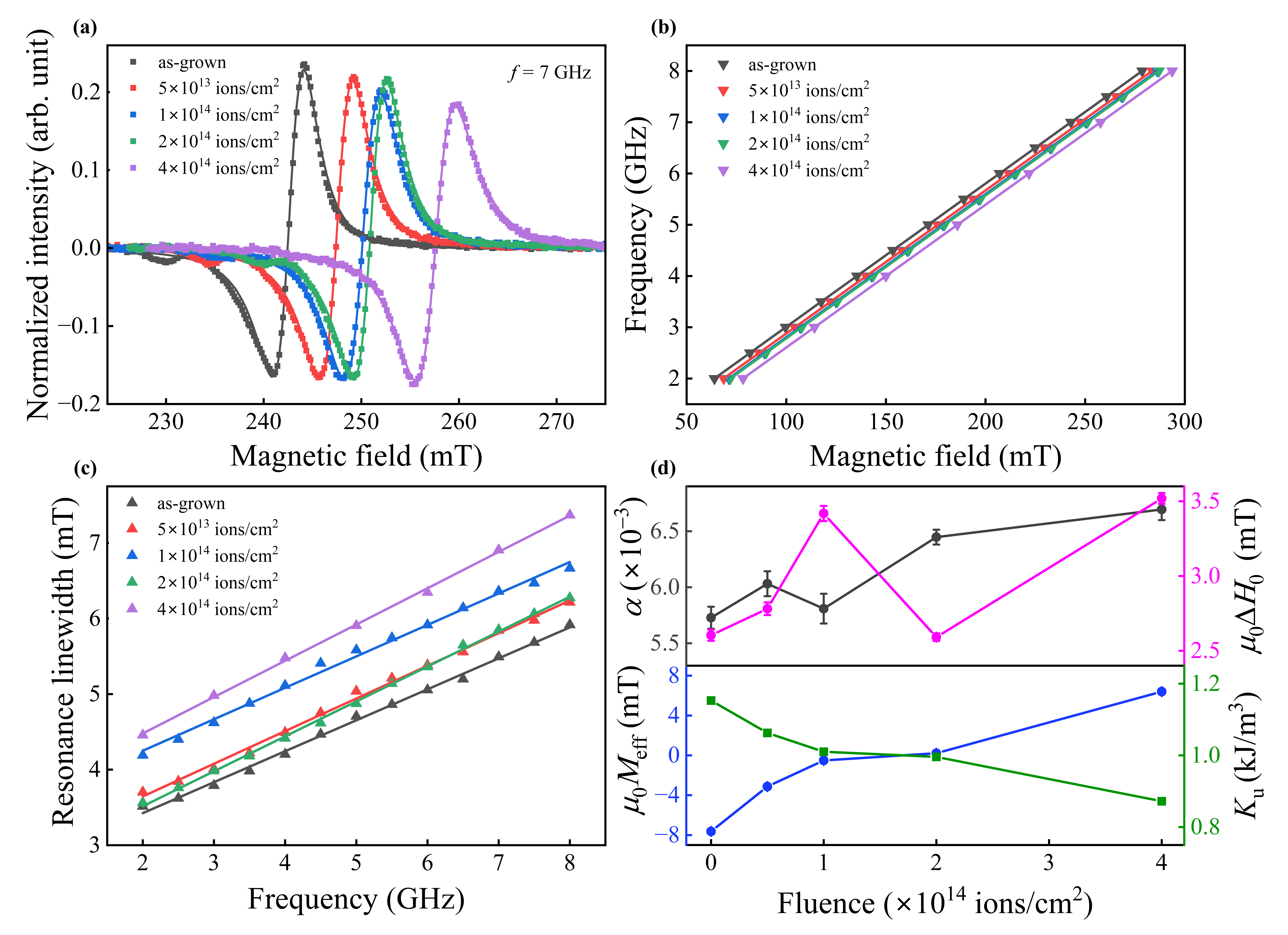}
    \caption{(a) FMR spectra recorded at 7 GHz on the as-grown Bi-YIG film and after He\(^+\) ion irradiation
    with different fluences. The symbols represent experimental data, while the lines are fits using the derivative of a Lorentzian function. (b) FMR frequency versus the magnetic resonance field for the Bi-YIG film after irradiation with different ion fluences. The lines represent Kittel formula fits to the experimental data. (c) Frequency dependence of the FMR linewidth for different ion fluences. Linear fits to the experimental data are used to extract the Gilbert damping parameter and the inhomogeneous broadening of the FMR linewidth. (d) Extracted values of $\alpha$, $\Delta H_\mathrm{0}$, $M_\mathrm{eff}$, and $K_\mathrm{u}$ as a function of irradiation fluence.}
    \label{fmr}
\end{figure*}

Bi-YIG films with a thickness of 20 nm were grown on  $5 \times 5$ mm (Mg,Zr)-substituted GGG(111) substrates using pulsed laser deposition (PLD) from a stoichiometric Bi$_{1}$Y$_{2}$Fe$_{5}$O$_{12}$ target with a KrF laser ($\lambda =$ 248 nm). The films were deposited at 700$\degree$C in an oxygen pressure of 0.05 mbar. To enhance their crystallinity, the films underwent \textit{in-situ} annealing at 700$\degree$C in 100 mbar oxygen after deposition. The lattice mismatch between Bi-YIG and sGGG induces an in-plane tensile strain in the Bi-YIG films, which, combined with the negative magnetostriction ($\lambda_{111}$) of Bi-YIG, results in PMA\cite{Soumah2018, Liu2019, Lin2020, Jia2023, Das2023}. Additionally, growth-induced anisotropy, originating from the preferential incorporation of Bi on Y sites, likely contributes to the PMA in the Bi-YIG films\cite{Soumah2018}.

The Bi-YIG films were irradiated with He$^{+}$ ions at room temperature using a Helium-S system from Spin-Ion Technologies by sweeping a millimeter-sized beam over the sample to enhance uniformity. The He$^{+}$ ion energy was set to 10 keV. Results from three films are shown. The first film was irradiated to achieve cumulative ion fluences at each stage of $5\times10^{13}$ ions/cm$^2$, $1\times10^{14}$ ions/cm$^2$, $2\times10^{14}$ ions/cm$^2$, and $4\times10^{14}$ ions/cm$^2$. After each irradiation step, the magnetic properties of this film were characterized using broadband FMR spectroscopy and MOKE microscopy. The film was then measured by vibrating sample magnetometry (VSM) and x-ray diffraction (XRD) before being analyzed with transmission electron microscopy (TEM). The results from this film are shown in Figs. 1-4 and the Supplemental Material. The second film was irradiated to a cumulative dose of $6\times10^{14}$ ions/cm$^2$ and then was used in the annealing experiments shown in Fig.\ 5. For comparison, an unirradiated film was also analyzed using VSM, XRD and TEM.

Figure \ref{moke2} displays polar MOKE hysteresis loops of the Bi-YIG film in its as-grown state and after irradiation with varying ion fluences. The as-grown Bi-YIG film shows an abrupt jump of the perpendicular magnetization around zero magnetic field, followed by the saturation at low magnetic fields. He\(^+\) ion irradiation gradually transforms the shape of the hysteresis curve to one which is slanted and closed. At higher ion fluences, the slope of the hysteresis curves decreases and the out-of-plane saturation field increases, suggesting a reduction in PMA.

VSM hysteresis loops recorded at room temperature with an in-plane magnetic field (Fig.\ \ref{moke2}(b)) confirm that the magnetization of the as-grown film at zero magnetic field is oriented out-of-plane, while the magnetization of the film with the highest ion fluence is oriented in-plane. The VSM measurements also reveal that the saturation magnetization, $M_\mathrm{s}=40\pm1.5$ kA/m, is unaffected by He$^+$ ion irradiation. The temperature dependence of \(M_\mathrm{s}\) for the irradiated film is fitted by the Bloch law\cite{Kuzmin2020} in Fig. \ref{moke2}(c). From this fit, we extract an exchange stiffness $A_\mathrm{ex}=0.79$ pJ/m. The data for the as-grown Bi-YIG film falls on the same fitting curve, demonstrating that the exchange stiffness is also not altered by ion irradiation.

\begin{figure}[bth]
\centering
    \includegraphics[width=1.0\linewidth]{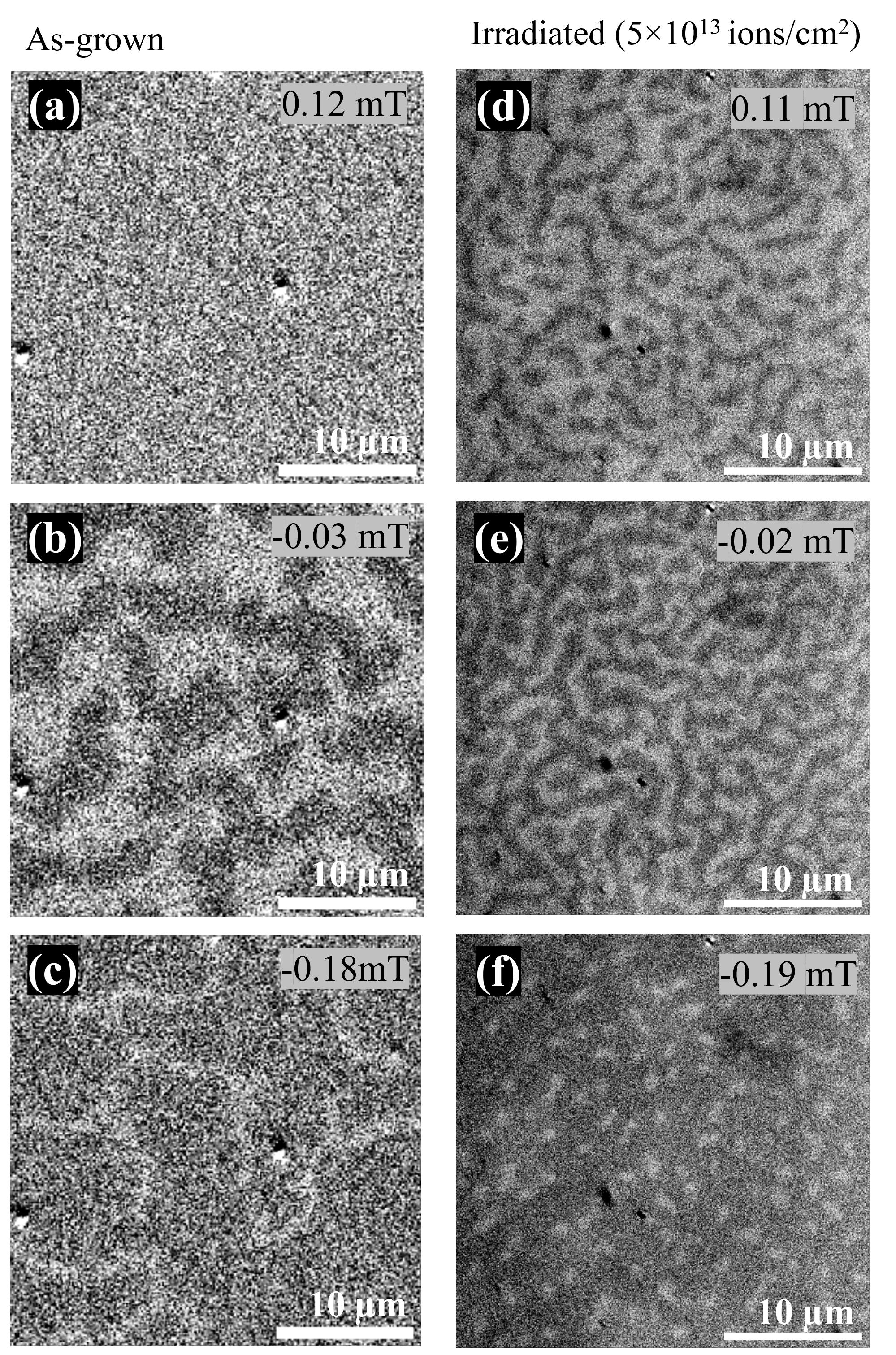}
    \caption{Polar MOKE microscopy images recorded during magnetization reversal in (a)-(c) the as-grown Bi-YIG film and (d)-(f) the Bi-YIG film after ion irradiation with a fluence of $5\times10^{13}$ ions/cm$^2$. The indicated applied out-of-plane fields are set after saturation in a positive field.}
    \label{sky}
\end{figure}

FMR spectra for the as-grown and ion irradiated Bi-YIG films, recorded at 7 GHz in a varying out-of-plane magnetic field, are shown in Fig.\ \ref{fmr}(a). The signal obtained by field modulation and lock-in amplification in the FMR spectrometer corresponds to the derivative of the microwave transmission coefficient (see Supplementary Note 1). He$^{+}$ ion irradiation shifts the FMR of the Bi-YIG film to higher magnetic fields, again indicating a decrease of PMA. We evaluate the dependence of various magnetic parameters on He$^{+}$ ion irradiation by fitting the FMR spectra recorded at different frequencies. Figures \ref{fmr}(b) and (c) summarize the magnetic resonance field ($H_\mathrm{res}$) and the FMR linewidth ($\Delta H$) extracted from the experimental data.       

The relation between the FMR frequency and the magnetic resonance field (Fig.\ \ref{fmr}(b)) is fitted using the Kittel formula for a perpendicularly magnetized film \cite{Gurevich1996}:
\begin{equation}
    f=\frac{\mu_0 \gamma}{2\pi}(H_\mathrm{res}- M_{\mathrm{eff}}). 
\end{equation}
Here, \(\mu_0\) is the vacuum permeability and \(\gamma\) is the gyromagnetic ratio. Fits to the data obtained for different ion fluences are shown as solid lines in Fig.\ \ref{fmr}(b). The extracted effective magnetization, $M_\mathrm{eff}$, is plotted as a function of ion fluence in Fig.\ \ref{fmr}(d). $M_\mathrm{eff}$ gradually increases and changes sign at a fluence of $2\times10^{14}$ ions/cm$^2$, indicating a reorientation of the film magnetization from out-of-plane to in-plane. This result is consistent with the changing shape of the MOKE and VSM hysteresis curves in Fig. \ref{moke2}. We estimate the strength of the uniaxial PMA, $K_\mathrm{u}$, using the experimentally derived values of $M_\mathrm{s}$ and $M_\mathrm{eff}$, and the relations\cite{Sbiaa2016}:
\begin{equation}
    H_\mathrm{Ku} = M_\mathrm{s} - M_\mathrm{eff} \hspace{0.5cm}\mathrm{and} \hspace{0.5cm}
    K_\mathrm{u} = \frac{\mu_{0}H_\mathrm{Ku}M_\mathrm{s}}{2}.\\
\end{equation}
Here, $H_\mathrm{Ku}$ is the magnetic anisotropy field. Since ion irradiation does not significantly change $M_\mathrm{s}$ (Fig. \ref{moke2}(b)), the increase and sign change of $M_\mathrm{eff}$ are due to a reduction in PMA, as shown in Fig.\ \ref{fmr}(d). The value of $K_\mathrm{u}$ decreases continuously with increasing ion fluence, and above $2\times10^{14}$ ions/cm$^2$, the magnetic shape anisotropy of the Bi-YIG film starts to dominate the PMA. Consequently, the film magnetization switches from out-of-plane to in-plane.  

We assess the effect of He$^{+}$ ion irradiation on the Gilbert damping parameter, $\alpha$, and the inhomogeneous broadening of the FMR linewidth, $\Delta H_0$, of the Bi-YIG film by fitting the experimental data to the following equation\cite{Nembach2011, Sbiaa2016}:
\begin{equation}
   \Delta H=\Delta H_0 + \frac{4\pi \alpha}{\gamma \mu_0}f.
\end{equation}
Both $\alpha$ and $\Delta H_\mathrm{0}$ exhibit a weak dependence on ion fluence, as illustrated by the extracted data in Fig. \ref{fmr}(d). For a fluence of $4\times10^{14}$ ions/cm$^2$, $\alpha$ is 14\% and $\Delta H_\mathrm{0}$ is 26\% larger than the values of the as-grown Bi-YIG film.

Figure \ref{sky} illustrates the effect of He\(^+\) ion irradiation on the magnetic domain structure of the Bi-YIG film. The MOKE microscopy images in Fig. \ref{sky}(a)-(c) show that magnetization reversal under applied out-of-plane magnetic fields in the as-grown Bi-YIG film occurs through the nucleation of magnetic stripe domains. Starting from positive saturation (Fig.\ \ref{sky}(a)), stripe domains emerge around zero applied magnetic field (Fig.\ \ref{sky}(b)) and then become thinner as the field is increased towards negative saturation (Fig.\ \ref{sky}(c)). After ion irradiation with a $5\times 10^{13}$ ions/cm$^2$, the formation of stripe domains is accompanied by the creation of magnetic bubbles, occurring before zero magnetic field when coming from positive saturation (Fig.\ \ref{sky}(d)). Near zero magnetic field, a dense stripe domain state is observed (Fig.\ \ref{sky}(e)), followed by a bubble domain state just below the negative saturation field (Fig.\ \ref{sky}(f)). The average diameter of the bubble domains is 1.2 \(\upmu\)m (see Supplementary Note 2). Additionally, the width of the stripe domains in the irradiated Bi-YIG film is smaller than in the as-grown film. Since the saturation magnetization remains unchanged by ion irradiation, the narrower stripe domains and the presence of magnetic bubbles in the irradiated Bi-YIG film are attributed to an ion-induced reduction of PMA (Fig. \ref{fmr}(d)). Bubble domains form due to the balance between the demagnetizing energy, which is reduced by the formation of domains, and the domain wall energy. In the absence of DMI, the domain wall energy is positive but decreases with reducing anisotropy, making bubble domains more stable.

The chirality of the magnetic bubble domains depends on the DMI strength. To estimate the DMI in our Bi-YIG film, both in the as-grown state and after ion irradiation with a fluence of $5\times 10^{13}$ ions/cm$^2$, we matched the width of the magnetic stripe domains in micromagnetic simulations to the MOKE microscopy data using the DMI as a fitting parameter (see Supplementary Note 3). The results indicate that both the as-grown and the ion irradiated Bi-YIG film exhibit a non-zero DMI constant of 0.006 mJ/m\(^2\). This DMI value is comparable to the reported DMI in previous studies on chiral domain walls in rare-earth iron garnets\cite{Velez2019, Ding2019, Caretta2020, Velez2022}. Our micromagnetic simulations suggest that the estimated DMI is insufficient to stabilize N\'{e}el-type skyrmions (see Supplementary Note 4). 

\begin{figure*}[htp]
\centering
    \includegraphics[width=1.0\linewidth]{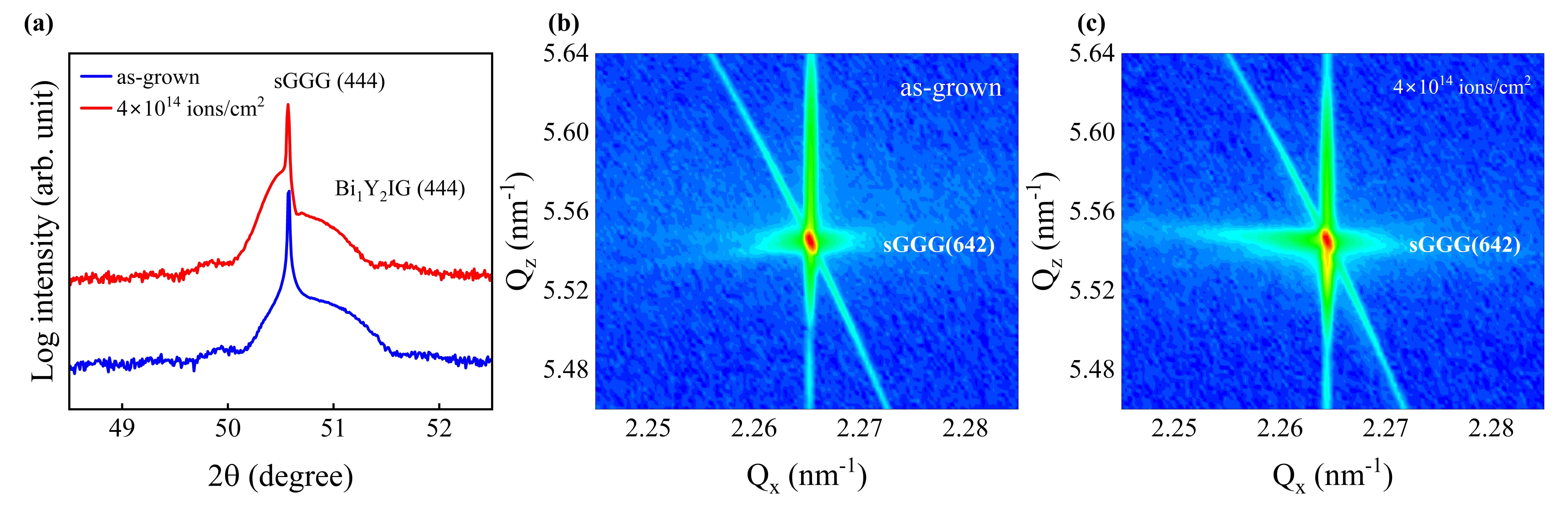}
    \caption{(a) XRD \(\uptheta\)-\(2\uptheta\) scans of an as-grown Bi-YIG film and a Bi-YIG film after ion irradiation with a fluence of $4\times10^{14}$ ions/cm$^2$. The scans are vertically offset for clarity. (b) XRD reciprocal space map of the sGGG(642) reflection recorded on the Bi-YIG film in the as-grown state. (c) XRD reciprocal space map of the sGGG(642) reflection recorded on the Bi-YIG film after irradiation with a fluence of $4\times10^{14}$ ions/cm$^2$.}
    \label{xrd-rsm}
\end{figure*}

The experimental data presented in Figs. 1 - 3 demonstrate that He$^+$ ion irradiation of the epitaxial Bi-YIG film on sGGG reduces the PMA, while other magnetic parameters either change minimally or remain constant. To investigate whether structural changes are responsible for the decrease in PMA, we conducted x-ray diffraction (XRD) measurements. Figure \ref{xrd-rsm}(a) compares XRD scans of an as-grown Bi-YIG film and a film irradiated with a fluence of $4\times10^{14}$ ions/cm$^2$. In the irradiated film, the Bi-YIG (444) peak (shoulder around 51$^\circ$) shifts to a lower $2\uptheta$ angle compared to the as-grown film, indicating an ion-induced expansion of the out-of-plane lattice constant and relaxation of the in-plane tensile strain. Additionally, a new diffraction shoulder appears on the left side of the sGGG (444) peak in the XRD scan post-irradiation. This new feature suggests an ion-induced expansion of the out-of-plane lattice parameter of a portion of the sGGG substrate. The intensity of this substrate shoulder peak is greater than that of the film peak (note the logarithmic scale), suggesting that the affected region in the sGGG substrate is thicker than the Bi-YIG film. This observation is consistent with stopping and range of ions in matter (SRIM) simulations\cite{ziegler1985}, which indicate that the stopping probability peaks around a depth of 100 nm for 10 keV He\(^+\) ions in Bi-YIG/sGGG (see Supplementary Note 5). The simulations further reveal that He\(^+\) ion irradiation primarily results in the formation of oxygen vacancies in both the Bi-YIG film and the sGGG substrate. 

\begin{figure*}[bt]
\centering
    \includegraphics[width=1.0\linewidth]{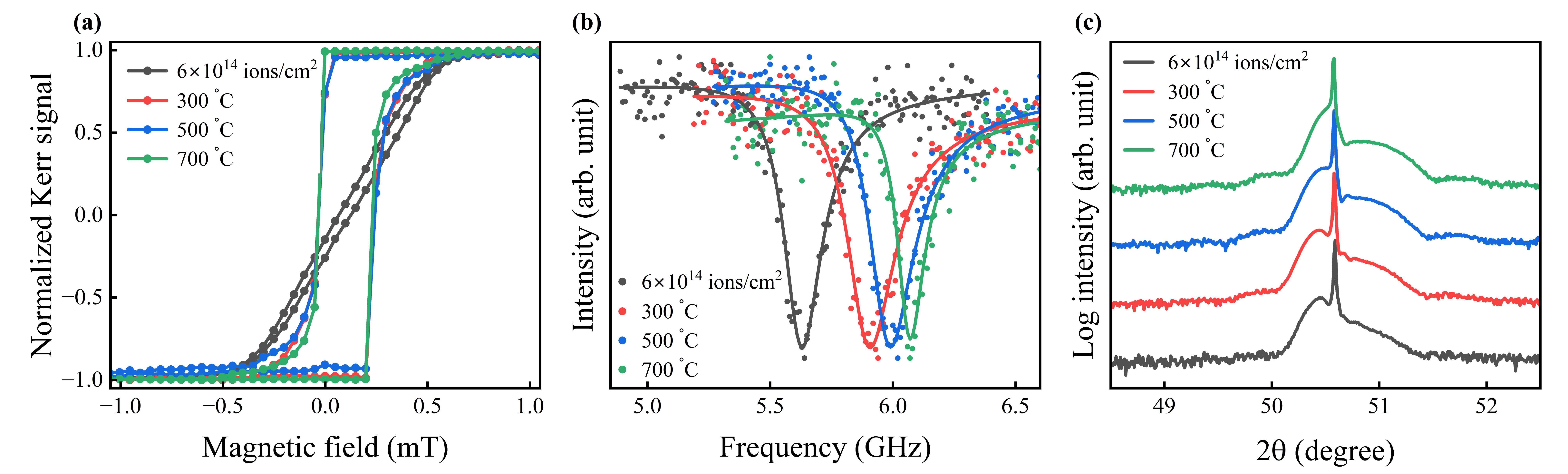}
    \caption{(a) Polar MOKE hysteresis loops of a Bi-YIG film after ion irradiation with a fluence of $6\times10^{14}$ ions/cm$^2$ and the same Bi-YIG film after thermal annealing at 300\(^\circ\)C, 500\(^\circ\)C, and 700\(^\circ\)C for 15 minutes in 0.05 mbar oxygen. (b) FMR spectra recorded at 205 mT for a Bi-YIG film after ion irradiation with a fluence of $6\times10^{14}$ ions/cm$^2$ and the same Bi-YIG film after thermal annealing at 300\(^\circ\)C, 500\(^\circ\)C, and 700\(^\circ\)C for 15 minutes in 0.05 mbar oxygen. (c) XRD \(\uptheta\)-\(2\uptheta\) scans of a Bi-YIG film after ion irradiation with a fluence of $6\times10^{14}$ ions/cm$^2$ and the same Bi-YIG film after thermal annealing at 300\(^\circ\)C, 500\(^\circ\)C, and 700\(^\circ\)C for 15 minutes in 0.05 mbar oxygen.}
    \label{annealing}
\end{figure*}

Figures \ref{xrd-rsm}(b) and (c) show XRD reciprocal space maps (RSMs) of an as-grown Bi-YIG film and a film after ion irradiation with a fluence of $4\times10^{14}$ ions/cm$^2$. Here, we observe a broadening of the sGGG diffraction peak towards lower $Q_\mathrm{z}$. Additionally, the sGGG reflection broadens slightly towards lower $Q_\mathrm{x}$. Despite this, the main sGGG peak and Bi-YIG film peak are observed at the same in-plane reciprocal lattice vector in both the as-grown state and after He$^{+}$ ion irradiation, indicating that good coherency between the crystal lattices of the sGGG substrate and the Bi-YIG film is maintained. The high structural quality of the Bi-YIG film on sGGG in both states is verified by transmission electron microscopy (TEM) (see Supplementary Note 6). 

We also investigated whether the ion-induced magnetic and structural changes in the Bi-YIG film could be reverted by thermal annealing. The annealing experiments were conducted on a Bi-YIG film irradiated with a fluence of $6\times10^{14}$ ions/cm$^2$. This film had a higher PMA after PLD growth than the one studied in Figs. 1-4, leading to a squarer hysteresis loop after growth, but similarly to the Bi-YIG film irradiated with $4\times10^{14}$ ions/cm$^2$, this film showed a reduction of PMA and a magnetization reorientation from out-of-plane to in-plane after ion irradiation. The irradiated sample was annealed in 0.05 mbar oxygen for 15 minutes at 300\(^\circ\)C, 500\(^\circ\)C, and 700\(^\circ\)C, with magnetic and structural measurements performed after each annealing step. Figure \ref{annealing} summarizes the effect of thermal annealing on polar MOKE hysteresis loops, FMR spectra, and XRD \(\uptheta\)-\(2\uptheta\) scans. Annealing the Bi-YIG film at 300\(^\circ\)C changes the shape of the polar hysteresis curve from slanted and closed to square and open. Thus, the magnetization of the irradiated Bi-YIG film switches back from in-plane to out-of-plane during annealing in oxygen, suggesting an enhancement of PMA. Further annealing at higher temperatures does not significantly alter the hysteresis curve. The upshift of the FMR frequency after thermal annealing at 300\(^\circ\)C confirms that the magnetic anisotropy increases (Fig. \ref{annealing}(b)). We note that the FMR data in these experiments were recorded with a vector network analyzer in a different FMR spectrometer than the one used to obtain the data shown in Fig. \ref{fmr}(a). The FMR spectra reveal that the PMA increases most after annealing at 300\(^\circ\)C, with further annealing at 500\(^\circ\)C and 700\(^\circ\)C slightly enhancing PMA. Analysis of the data shows that He$^{+}$ ion irradiation reduces $K_\mathrm{u}$ by 27\(\%\), while thermal annealing increases it by 23\(\%\). Thus, most of the PMA from the as-grown state is recovered through annealing. The XRD measurements in Fig. \ref{annealing}(c) support this finding, as thermal annealing reduces the diffraction shoulder on the left side of the sGGG peak and the Bi-YIG film peak shifts to a higher \(2\uptheta\) angle. From these results, we conclude that the He$^{+}$-ion-induced changes in the sGGG substrate and the Bi-YIG film are largely reversed by thermal annealing.

Taking all data into account, the following picture emerges: He$^{+}$ ion irradiation of the Bi-YIG film produces oxygen vacancies, known to expand the crystal lattice of YIG\cite{Dumont2007,Noun2010}. This expansion reduces the lattice mismatch between the Bi-YIG film and the sGGG substrate, leading to the relaxation of tensile strain in the Bi-YIG film and a decrease in its strain-induced PMA. Ion irradiation also creates vacancies in the sGGG substrate, as indicated by the appearance of a diffraction shoulder in XRD measurements and SRIM simulations. However, since vacancy formation in the sGGG substrate predominately alters its out-of-plane lattice parameter (see RSM in Fig. \ref{xrd-rsm}(c)), the structural changes in the substrate do not significantly alter the strain in the Bi-YIG film. Annealing in an oxygen atmosphere reduces the number of oxygen vacancies in both sGGG and Bi-YIG, restoring most of the tensile strain and PMA in the film.

In summary, we investigated the effects of He$^+$ ion irradiation on the magnetic and structural properties of epitaxial Bi-YIG films on sGGG substrates. In the as-grown state, the Bi-YIG films exhibit strong PMA. He$^+$ ion irradiation reduces the tensile strain and PMA of the Bi-YIG films via the formation of oxygen vacancies, causing the magnetization to switch from out-of-plane to in-plane at a fluence of $2\times10^{14}$ ions/cm$^2$. Other magnetic parameters, such as saturation magnetization, magnetic damping parameter, and inhomogeneous broadening of the FMR linewidth, are minimally affected or unchanged by ion irradiation. The DMI also remains constant. The distinction between a changing PMA and constant DMI suggests that the DMI in this system is interfacial in nature, dependent on the non-varying in-plane lattice parameter, while a volume magnetoelastic effect induces the PMA. The combination of reducing PMA and constant DMI results in the formation of bubble domains at a fluence of $5\times10^{13}$ ions/cm$^2$. Due to the small DMI, these magnetic bubbles are not identified as N\'{e}el skyrmions. Thermal annealing in an oxygen atmosphere reverses most of the effects of He$^+$ ion irradiation by removing oxygen vacancies. He$^+$ ion irradiation of garnet films, as demonstrated for Bi-YIG films in this study, enables reliable tuning of magnetic anisotropy while minimally affecting other magnetic parameters. 

\begin{acknowledgments}
This project has received funding from the European Union’s Horizon 2020 research and innovation program under the Marie Skłodowska-Curie grant agreement No. 860060 “Magnetism and the effects of Electric Field” (MagnEFi). This work was supported by the Academy of Finland (grant no. 338748). We acknowledge the x-ray and transmission electron microscopy facilities at the OtaNano-Nanomicroscopy Center at Aalto University, and the computational resources provided by the Aalto Science-IT project.
\end{acknowledgments}

\bibliography{paper2}

\begin{thebibliography}{47}%
\makeatletter
\providecommand \@ifxundefined [1]{%
 \@ifx{#1\undefined}
}%
\providecommand \@ifnum [1]{%
 \ifnum #1\expandafter \@firstoftwo
 \else \expandafter \@secondoftwo
 \fi
}%
\providecommand \@ifx [1]{%
 \ifx #1\expandafter \@firstoftwo
 \else \expandafter \@secondoftwo
 \fi
}%
\providecommand \natexlab [1]{#1}%
\providecommand \enquote  [1]{``#1''}%
\providecommand \bibnamefont  [1]{#1}%
\providecommand \bibfnamefont [1]{#1}%
\providecommand \citenamefont [1]{#1}%
\providecommand \href@noop [0]{\@secondoftwo}%
\providecommand \href [0]{\begingroup \@sanitize@url \@href}%
\providecommand \@href[1]{\@@startlink{#1}\@@href}%
\providecommand \@@href[1]{\endgroup#1\@@endlink}%
\providecommand \@sanitize@url [0]{\catcode `\\12\catcode `\$12\catcode
  `\&12\catcode `\#12\catcode `\^12\catcode `\_12\catcode `\%12\relax}%
\providecommand \@@startlink[1]{}%
\providecommand \@@endlink[0]{}%
\providecommand \url  [0]{\begingroup\@sanitize@url \@url }%
\providecommand \@url [1]{\endgroup\@href {#1}{\urlprefix }}%
\providecommand \urlprefix  [0]{URL }%
\providecommand \Eprint [0]{\href }%
\providecommand \doibase [0]{http://dx.doi.org/}%
\providecommand \selectlanguage [0]{\@gobble}%
\providecommand \bibinfo  [0]{\@secondoftwo}%
\providecommand \bibfield  [0]{\@secondoftwo}%
\providecommand \translation [1]{[#1]}%
\providecommand \BibitemOpen [0]{}%
\providecommand \bibitemStop [0]{}%
\providecommand \bibitemNoStop [0]{.\EOS\space}%
\providecommand \EOS [0]{\spacefactor3000\relax}%
\providecommand \BibitemShut  [1]{\csname bibitem#1\endcsname}%
\let\auto@bib@innerbib\@empty
\bibitem [{\citenamefont {Ding}\ \emph {et~al.}(2020)\citenamefont {Ding},
  \citenamefont {Liu}, \citenamefont {Zhang}, \citenamefont {Erugu},
  \citenamefont {Quan}, \citenamefont {Yu}, \citenamefont {McCollum},
  \citenamefont {Mo}, \citenamefont {Yang}, \citenamefont {Ding}, \citenamefont
  {Xu}, \citenamefont {Tang}, \citenamefont {Yang},\ and\ \citenamefont
  {Wu}}]{Ding2020}%
  \BibitemOpen
  \bibfield  {author} {\bibinfo {author} {\bibfnamefont {J.}~\bibnamefont
  {Ding}}, \bibinfo {author} {\bibfnamefont {C.}~\bibnamefont {Liu}}, \bibinfo
  {author} {\bibfnamefont {Y.}~\bibnamefont {Zhang}}, \bibinfo {author}
  {\bibfnamefont {U.}~\bibnamefont {Erugu}}, \bibinfo {author} {\bibfnamefont
  {Z.}~\bibnamefont {Quan}}, \bibinfo {author} {\bibfnamefont {R.}~\bibnamefont
  {Yu}}, \bibinfo {author} {\bibfnamefont {E.}~\bibnamefont {McCollum}},
  \bibinfo {author} {\bibfnamefont {S.}~\bibnamefont {Mo}}, \bibinfo {author}
  {\bibfnamefont {S.}~\bibnamefont {Yang}}, \bibinfo {author} {\bibfnamefont
  {H.}~\bibnamefont {Ding}}, \bibinfo {author} {\bibfnamefont {X.}~\bibnamefont
  {Xu}}, \bibinfo {author} {\bibfnamefont {J.}~\bibnamefont {Tang}}, \bibinfo
  {author} {\bibfnamefont {X.}~\bibnamefont {Yang}}, \ and\ \bibinfo {author}
  {\bibfnamefont {M.}~\bibnamefont {Wu}},\ }\bibfield  {title} {\enquote
  {\bibinfo {title} {Nanometer-thick yttrium iron garnet films with
  perpendicular anisotropy and low damping},}\ }\href {\doibase
  10.1103/physrevapplied.14.014017} {\bibfield  {journal} {\bibinfo  {journal}
  {Physical Review Applied}\ }\textbf {\bibinfo {volume} {14}},\ \bibinfo
  {pages} {014017} (\bibinfo {year} {2020})}\BibitemShut {NoStop}%
\bibitem [{\citenamefont {Fu}\ \emph {et~al.}(2017)\citenamefont {Fu},
  \citenamefont {Hua}, \citenamefont {Wen}, \citenamefont {Xue}, \citenamefont
  {Ding}, \citenamefont {Wang}, \citenamefont {Yu}, \citenamefont {Liu},
  \citenamefont {Han}, \citenamefont {Wang}, \citenamefont {Du}, \citenamefont
  {Yang},\ and\ \citenamefont {Yang}}]{Fu2017}%
  \BibitemOpen
  \bibfield  {author} {\bibinfo {author} {\bibfnamefont {J.}~\bibnamefont
  {Fu}}, \bibinfo {author} {\bibfnamefont {M.}~\bibnamefont {Hua}}, \bibinfo
  {author} {\bibfnamefont {X.}~\bibnamefont {Wen}}, \bibinfo {author}
  {\bibfnamefont {M.}~\bibnamefont {Xue}}, \bibinfo {author} {\bibfnamefont
  {S.}~\bibnamefont {Ding}}, \bibinfo {author} {\bibfnamefont {M.}~\bibnamefont
  {Wang}}, \bibinfo {author} {\bibfnamefont {P.}~\bibnamefont {Yu}}, \bibinfo
  {author} {\bibfnamefont {S.}~\bibnamefont {Liu}}, \bibinfo {author}
  {\bibfnamefont {J.}~\bibnamefont {Han}}, \bibinfo {author} {\bibfnamefont
  {C.}~\bibnamefont {Wang}}, \bibinfo {author} {\bibfnamefont {H.}~\bibnamefont
  {Du}}, \bibinfo {author} {\bibfnamefont {Y.}~\bibnamefont {Yang}}, \ and\
  \bibinfo {author} {\bibfnamefont {J.}~\bibnamefont {Yang}},\ }\bibfield
  {title} {\enquote {\bibinfo {title} {Epitaxial growth of
  {Y}$_{3}${Fe}$_{5}${O}$_{12}$ thin films with perpendicular magnetic
  anisotropy},}\ }\href {\doibase 10.1063/1.4983783} {\bibfield  {journal}
  {\bibinfo  {journal} {Applied Physics Letters}\ }\textbf {\bibinfo {volume}
  {110}},\ \bibinfo {pages} {202403} (\bibinfo {year} {2017})}\BibitemShut
  {NoStop}%
\bibitem [{\citenamefont {Chen}\ \emph {et~al.}(2022)\citenamefont {Chen},
  \citenamefont {Xie}, \citenamefont {Yang}, \citenamefont {Gao}, \citenamefont
  {Liu}, \citenamefont {Qin}, \citenamefont {Yan}, \citenamefont {Tan},
  \citenamefont {Chen}, \citenamefont {Gong}, \citenamefont {Li}, \citenamefont
  {Bi}, \citenamefont {Liu},\ and\ \citenamefont {Deng}}]{Chen2022}%
  \BibitemOpen
  \bibfield  {author} {\bibinfo {author} {\bibfnamefont {S.}~\bibnamefont
  {Chen}}, \bibinfo {author} {\bibfnamefont {Y.}~\bibnamefont {Xie}}, \bibinfo
  {author} {\bibfnamefont {Y.}~\bibnamefont {Yang}}, \bibinfo {author}
  {\bibfnamefont {D.}~\bibnamefont {Gao}}, \bibinfo {author} {\bibfnamefont
  {D.}~\bibnamefont {Liu}}, \bibinfo {author} {\bibfnamefont {L.}~\bibnamefont
  {Qin}}, \bibinfo {author} {\bibfnamefont {W.}~\bibnamefont {Yan}}, \bibinfo
  {author} {\bibfnamefont {B.}~\bibnamefont {Tan}}, \bibinfo {author}
  {\bibfnamefont {Q.}~\bibnamefont {Chen}}, \bibinfo {author} {\bibfnamefont
  {T.}~\bibnamefont {Gong}}, \bibinfo {author} {\bibfnamefont {E.}~\bibnamefont
  {Li}}, \bibinfo {author} {\bibfnamefont {L.}~\bibnamefont {Bi}}, \bibinfo
  {author} {\bibfnamefont {T.}~\bibnamefont {Liu}}, \ and\ \bibinfo {author}
  {\bibfnamefont {L.}~\bibnamefont {Deng}},\ }\bibfield  {title} {\enquote
  {\bibinfo {title} {The 50 nm-thick yttrium iron garnet films with
  perpendicular magnetic anisotropy},}\ }\href {\doibase
  10.1088/1674-1056/ac4cc4} {\bibfield  {journal} {\bibinfo  {journal} {Chinese
  Physics B}\ }\textbf {\bibinfo {volume} {31}},\ \bibinfo {pages} {048503}
  (\bibinfo {year} {2022})}\BibitemShut {NoStop}%
\bibitem [{\citenamefont {Wang}\ \emph {et~al.}(2017)\citenamefont {Wang},
  \citenamefont {Liang}, \citenamefont {Zhang}, \citenamefont {Liang},
  \citenamefont {Zhu}, \citenamefont {Qin}, \citenamefont {Gao}, \citenamefont
  {Peng}, \citenamefont {Sun},\ and\ \citenamefont {Bi}}]{Wang2017a}%
  \BibitemOpen
  \bibfield  {author} {\bibinfo {author} {\bibfnamefont {C.~T.}\ \bibnamefont
  {Wang}}, \bibinfo {author} {\bibfnamefont {X.~F.}\ \bibnamefont {Liang}},
  \bibinfo {author} {\bibfnamefont {Y.}~\bibnamefont {Zhang}}, \bibinfo
  {author} {\bibfnamefont {X.}~\bibnamefont {Liang}}, \bibinfo {author}
  {\bibfnamefont {Y.~P.}\ \bibnamefont {Zhu}}, \bibinfo {author} {\bibfnamefont
  {J.}~\bibnamefont {Qin}}, \bibinfo {author} {\bibfnamefont {Y.}~\bibnamefont
  {Gao}}, \bibinfo {author} {\bibfnamefont {B.}~\bibnamefont {Peng}}, \bibinfo
  {author} {\bibfnamefont {N.~X.}\ \bibnamefont {Sun}}, \ and\ \bibinfo
  {author} {\bibfnamefont {L.}~\bibnamefont {Bi}},\ }\bibfield  {title}
  {\enquote {\bibinfo {title} {Controlling the magnetic anisotropy in epitaxial
  {Y}$_{3}${Fe}$_{5}${O}$_{12}$ films by manganese doping},}\ }\href@noop {}
  {\bibfield  {journal} {\bibinfo  {journal} {Phys. Rev. B}\ }\textbf {\bibinfo
  {volume} {96}},\ \bibinfo {pages} {224403} (\bibinfo {year}
  {2017})}\BibitemShut {NoStop}%
\bibitem [{\citenamefont {Lage}\ \emph {et~al.}(2017)\citenamefont {Lage},
  \citenamefont {Beran}, \citenamefont {Quindeau}, \citenamefont {Ohnoutek},
  \citenamefont {Kucera}, \citenamefont {Antos}, \citenamefont {Sani},
  \citenamefont {Dionne}, \citenamefont {Veis},\ and\ \citenamefont
  {Ross}}]{Lage2017}%
  \BibitemOpen
  \bibfield  {author} {\bibinfo {author} {\bibfnamefont {E.}~\bibnamefont
  {Lage}}, \bibinfo {author} {\bibfnamefont {L.}~\bibnamefont {Beran}},
  \bibinfo {author} {\bibfnamefont {A.~U.}\ \bibnamefont {Quindeau}}, \bibinfo
  {author} {\bibfnamefont {L.}~\bibnamefont {Ohnoutek}}, \bibinfo {author}
  {\bibfnamefont {M.}~\bibnamefont {Kucera}}, \bibinfo {author} {\bibfnamefont
  {R.}~\bibnamefont {Antos}}, \bibinfo {author} {\bibfnamefont {S.~R.}\
  \bibnamefont {Sani}}, \bibinfo {author} {\bibfnamefont {G.~F.}\ \bibnamefont
  {Dionne}}, \bibinfo {author} {\bibfnamefont {M.}~\bibnamefont {Veis}}, \ and\
  \bibinfo {author} {\bibfnamefont {C.~A.}\ \bibnamefont {Ross}},\ }\bibfield
  {title} {\enquote {\bibinfo {title} {Temperature-dependent {F}araday rotation
  and magnetization reorientation in cerium-substituted yttrium iron garnet
  thin films},}\ }\href@noop {} {\bibfield  {journal} {\bibinfo  {journal}
  {{APL} Materials}\ }\textbf {\bibinfo {volume} {5}},\ \bibinfo {pages}
  {036104} (\bibinfo {year} {2017})}\BibitemShut {NoStop}%
\bibitem [{\citenamefont {Soumah}\ \emph {et~al.}(2018)\citenamefont {Soumah},
  \citenamefont {Beaulieu}, \citenamefont {Qassym}, \citenamefont
  {Carr{\'{e}}t{\'{e}}ro}, \citenamefont {Jacquet}, \citenamefont
  {Lebourgeois}, \citenamefont {Youssef}, \citenamefont {Bortolotti},
  \citenamefont {Cros},\ and\ \citenamefont {Anane}}]{Soumah2018}%
  \BibitemOpen
  \bibfield  {author} {\bibinfo {author} {\bibfnamefont {L.}~\bibnamefont
  {Soumah}}, \bibinfo {author} {\bibfnamefont {N.}~\bibnamefont {Beaulieu}},
  \bibinfo {author} {\bibfnamefont {L.}~\bibnamefont {Qassym}}, \bibinfo
  {author} {\bibfnamefont {C.}~\bibnamefont {Carr{\'{e}}t{\'{e}}ro}}, \bibinfo
  {author} {\bibfnamefont {E.}~\bibnamefont {Jacquet}}, \bibinfo {author}
  {\bibfnamefont {R.}~\bibnamefont {Lebourgeois}}, \bibinfo {author}
  {\bibfnamefont {J.~B.}\ \bibnamefont {Youssef}}, \bibinfo {author}
  {\bibfnamefont {P.}~\bibnamefont {Bortolotti}}, \bibinfo {author}
  {\bibfnamefont {V.}~\bibnamefont {Cros}}, \ and\ \bibinfo {author}
  {\bibfnamefont {A.}~\bibnamefont {Anane}},\ }\bibfield  {title} {\enquote
  {\bibinfo {title} {Ultra-low damping insulating magnetic thin films get
  perpendicular},}\ }\href@noop {} {\bibfield  {journal} {\bibinfo  {journal}
  {Nature Communications}\ }\textbf {\bibinfo {volume} {9}},\ \bibinfo {pages}
  {3355} (\bibinfo {year} {2018})}\BibitemShut {NoStop}%
\bibitem [{\citenamefont {Liu}\ \emph {et~al.}(2019)\citenamefont {Liu},
  \citenamefont {Yang}, \citenamefont {Zhang}, \citenamefont {Wu},\ and\
  \citenamefont {Zhang}}]{Liu2019}%
  \BibitemOpen
  \bibfield  {author} {\bibinfo {author} {\bibfnamefont {X.}~\bibnamefont
  {Liu}}, \bibinfo {author} {\bibfnamefont {Q.}~\bibnamefont {Yang}}, \bibinfo
  {author} {\bibfnamefont {D.}~\bibnamefont {Zhang}}, \bibinfo {author}
  {\bibfnamefont {Y.}~\bibnamefont {Wu}}, \ and\ \bibinfo {author}
  {\bibfnamefont {H.}~\bibnamefont {Zhang}},\ }\bibfield  {title} {\enquote
  {\bibinfo {title} {Magnetic properties of bismuth substituted yttrium iron
  garnet film with perpendicular magnetic anisotropy},}\ }\href {\doibase
  10.1063/1.5122998} {\bibfield  {journal} {\bibinfo  {journal} {{AIP}
  Advances}\ }\textbf {\bibinfo {volume} {9}},\ \bibinfo {pages} {115001}
  (\bibinfo {year} {2019})}\BibitemShut {NoStop}%
\bibitem [{\citenamefont {Lin}\ \emph {et~al.}(2020)\citenamefont {Lin},
  \citenamefont {Jin}, \citenamefont {Zhang}, \citenamefont {Zhong},
  \citenamefont {Yang}, \citenamefont {Rao},\ and\ \citenamefont
  {Li}}]{Lin2020}%
  \BibitemOpen
  \bibfield  {author} {\bibinfo {author} {\bibfnamefont {Y.}~\bibnamefont
  {Lin}}, \bibinfo {author} {\bibfnamefont {L.}~\bibnamefont {Jin}}, \bibinfo
  {author} {\bibfnamefont {H.}~\bibnamefont {Zhang}}, \bibinfo {author}
  {\bibfnamefont {Z.}~\bibnamefont {Zhong}}, \bibinfo {author} {\bibfnamefont
  {Q.}~\bibnamefont {Yang}}, \bibinfo {author} {\bibfnamefont {Y.}~\bibnamefont
  {Rao}}, \ and\ \bibinfo {author} {\bibfnamefont {M.}~\bibnamefont {Li}},\
  }\bibfield  {title} {\enquote {\bibinfo {title} {Bi-{YIG} ferrimagnetic
  insulator nanometer films with large perpendicular magnetic anisotropy and
  narrow ferromagnetic resonance linewidth},}\ }\href@noop {} {\bibfield
  {journal} {\bibinfo  {journal} {Journal of Magnetism and Magnetic Materials}\
  }\textbf {\bibinfo {volume} {496}},\ \bibinfo {pages} {165886} (\bibinfo
  {year} {2020})}\BibitemShut {NoStop}%
\bibitem [{\citenamefont {Kuila}\ \emph {et~al.}(2022)\citenamefont {Kuila},
  \citenamefont {Sagdeo}, \citenamefont {Longchar}, \citenamefont {Choudhary},
  \citenamefont {Srinath},\ and\ \citenamefont {Reddy}}]{Kuila2022}%
  \BibitemOpen
  \bibfield  {author} {\bibinfo {author} {\bibfnamefont {M.}~\bibnamefont
  {Kuila}}, \bibinfo {author} {\bibfnamefont {A.}~\bibnamefont {Sagdeo}},
  \bibinfo {author} {\bibfnamefont {L.~A.}\ \bibnamefont {Longchar}}, \bibinfo
  {author} {\bibfnamefont {R.~J.}\ \bibnamefont {Choudhary}}, \bibinfo {author}
  {\bibfnamefont {S.}~\bibnamefont {Srinath}}, \ and\ \bibinfo {author}
  {\bibfnamefont {V.~R.}\ \bibnamefont {Reddy}},\ }\bibfield  {title} {\enquote
  {\bibinfo {title} {Robust perpendicular magnetic anisotropy in {C}e
  substituted yttrium iron garnet epitaxial thin films},}\ }\href {\doibase
  10.1063/5.0085572} {\bibfield  {journal} {\bibinfo  {journal} {Journal of
  Applied Physics}\ }\textbf {\bibinfo {volume} {131}},\ \bibinfo {pages}
  {203901} (\bibinfo {year} {2022})}\BibitemShut {NoStop}%
\bibitem [{\citenamefont {Böttcher}\ \emph {et~al.}(2022)\citenamefont
  {Böttcher}, \citenamefont {Ruhwedel}, \citenamefont {Levchenko},
  \citenamefont {Wang}, \citenamefont {Chumak}, \citenamefont {Popov},
  \citenamefont {Zavislyak}, \citenamefont {Dubs}, \citenamefont {Surzhenko},
  \citenamefont {Hillebrands}, \citenamefont {Chumak},\ and\ \citenamefont
  {Pirro}}]{Boettcher2022}%
  \BibitemOpen
  \bibfield  {author} {\bibinfo {author} {\bibfnamefont {T.}~\bibnamefont
  {Böttcher}}, \bibinfo {author} {\bibfnamefont {M.}~\bibnamefont {Ruhwedel}},
  \bibinfo {author} {\bibfnamefont {K.~O.}\ \bibnamefont {Levchenko}}, \bibinfo
  {author} {\bibfnamefont {Q.}~\bibnamefont {Wang}}, \bibinfo {author}
  {\bibfnamefont {H.~L.}\ \bibnamefont {Chumak}}, \bibinfo {author}
  {\bibfnamefont {M.~A.}\ \bibnamefont {Popov}}, \bibinfo {author}
  {\bibfnamefont {I.~V.}\ \bibnamefont {Zavislyak}}, \bibinfo {author}
  {\bibfnamefont {C.}~\bibnamefont {Dubs}}, \bibinfo {author} {\bibfnamefont
  {O.}~\bibnamefont {Surzhenko}}, \bibinfo {author} {\bibfnamefont
  {B.}~\bibnamefont {Hillebrands}}, \bibinfo {author} {\bibfnamefont {A.~V.}\
  \bibnamefont {Chumak}}, \ and\ \bibinfo {author} {\bibfnamefont
  {P.}~\bibnamefont {Pirro}},\ }\bibfield  {title} {\enquote {\bibinfo {title}
  {Fast long-wavelength exchange spin waves in partially compensated
  {G}a:{YIG}},}\ }\href {\doibase 10.1063/5.0082724} {\bibfield  {journal}
  {\bibinfo  {journal} {Applied Physics Letters}\ }\textbf {\bibinfo {volume}
  {120}},\ \bibinfo {pages} {102401} (\bibinfo {year} {2022})}\BibitemShut
  {NoStop}%
\bibitem [{\citenamefont {Jia}\ \emph {et~al.}(2023)\citenamefont {Jia},
  \citenamefont {Liang}, \citenamefont {Pan}, \citenamefont {Wang},
  \citenamefont {Lv}, \citenamefont {Yan}, \citenamefont {Jin}, \citenamefont
  {Hou}, \citenamefont {Wang},\ and\ \citenamefont {Wu}}]{Jia2023}%
  \BibitemOpen
  \bibfield  {author} {\bibinfo {author} {\bibfnamefont {Y.}~\bibnamefont
  {Jia}}, \bibinfo {author} {\bibfnamefont {Z.}~\bibnamefont {Liang}}, \bibinfo
  {author} {\bibfnamefont {H.}~\bibnamefont {Pan}}, \bibinfo {author}
  {\bibfnamefont {Q.}~\bibnamefont {Wang}}, \bibinfo {author} {\bibfnamefont
  {Q.}~\bibnamefont {Lv}}, \bibinfo {author} {\bibfnamefont {Y.}~\bibnamefont
  {Yan}}, \bibinfo {author} {\bibfnamefont {F.}~\bibnamefont {Jin}}, \bibinfo
  {author} {\bibfnamefont {D.}~\bibnamefont {Hou}}, \bibinfo {author}
  {\bibfnamefont {L.}~\bibnamefont {Wang}}, \ and\ \bibinfo {author}
  {\bibfnamefont {W.}~\bibnamefont {Wu}},\ }\bibfield  {title} {\enquote
  {\bibinfo {title} {Bismuth doping enhanced tunability of strain-controlled
  magnetic anisotropy in epitaxial $\mathrm{{Y}_{3}{Fe}_{5}{O}_{12}}$(111)
  films},}\ }\href {\doibase 10.1088/1674-1056/ac67cc} {\bibfield  {journal}
  {\bibinfo  {journal} {Chinese Physics B}\ }\textbf {\bibinfo {volume} {32}},\
  \bibinfo {pages} {027501} (\bibinfo {year} {2023})}\BibitemShut {NoStop}%
\bibitem [{\citenamefont {Das}\ \emph {et~al.}(2023)\citenamefont {Das},
  \citenamefont {Mansell}, \citenamefont {Flajšman}, \citenamefont {Yao},\
  and\ \citenamefont {van Dijken}}]{Das2023}%
  \BibitemOpen
  \bibfield  {author} {\bibinfo {author} {\bibfnamefont {S.}~\bibnamefont
  {Das}}, \bibinfo {author} {\bibfnamefont {R.}~\bibnamefont {Mansell}},
  \bibinfo {author} {\bibfnamefont {L.}~\bibnamefont {Flajšman}}, \bibinfo
  {author} {\bibfnamefont {L.}~\bibnamefont {Yao}}, \ and\ \bibinfo {author}
  {\bibfnamefont {S.}~\bibnamefont {van Dijken}},\ }\bibfield  {title}
  {\enquote {\bibinfo {title} {Perpendicular magnetic anisotropy in
  {B}i-substituted yttrium iron garnet films},}\ }\href@noop {} {\bibfield
  {journal} {\bibinfo  {journal} {Journal of Applied Physics}\ }\textbf
  {\bibinfo {volume} {134}} (\bibinfo {year} {2023})}\BibitemShut {NoStop}%
\bibitem [{\citenamefont {Wu}\ \emph {et~al.}(2018)\citenamefont {Wu},
  \citenamefont {Tseng}, \citenamefont {Fanchiang}, \citenamefont {Cheng},
  \citenamefont {Lin}, \citenamefont {Yeh}, \citenamefont {Yang}, \citenamefont
  {Wu}, \citenamefont {Liu}, \citenamefont {Wu}, \citenamefont {Hong},\ and\
  \citenamefont {Kwo}}]{Wu2018}%
  \BibitemOpen
  \bibfield  {author} {\bibinfo {author} {\bibfnamefont {C.~N.}\ \bibnamefont
  {Wu}}, \bibinfo {author} {\bibfnamefont {C.~C.}\ \bibnamefont {Tseng}},
  \bibinfo {author} {\bibfnamefont {Y.~T.}\ \bibnamefont {Fanchiang}}, \bibinfo
  {author} {\bibfnamefont {C.~K.}\ \bibnamefont {Cheng}}, \bibinfo {author}
  {\bibfnamefont {K.~Y.}\ \bibnamefont {Lin}}, \bibinfo {author} {\bibfnamefont
  {S.~L.}\ \bibnamefont {Yeh}}, \bibinfo {author} {\bibfnamefont {S.~R.}\
  \bibnamefont {Yang}}, \bibinfo {author} {\bibfnamefont {C.~T.}\ \bibnamefont
  {Wu}}, \bibinfo {author} {\bibfnamefont {T.}~\bibnamefont {Liu}}, \bibinfo
  {author} {\bibfnamefont {M.}~\bibnamefont {Wu}}, \bibinfo {author}
  {\bibfnamefont {M.}~\bibnamefont {Hong}}, \ and\ \bibinfo {author}
  {\bibfnamefont {J.}~\bibnamefont {Kwo}},\ }\bibfield  {title} {\enquote
  {\bibinfo {title} {High-quality thulium iron garnet films with tunable
  perpendicular magnetic anisotropy by off-axis sputtering {\textendash}
  correlation between magnetic properties and film strain},}\ }\href@noop {}
  {\bibfield  {journal} {\bibinfo  {journal} {Scientific Reports}\ }\textbf
  {\bibinfo {volume} {8}} (\bibinfo {year} {2018})}\BibitemShut {NoStop}%
\bibitem [{\citenamefont {Rosenberg}\ \emph {et~al.}(2018)\citenamefont
  {Rosenberg}, \citenamefont {Beran}, \citenamefont {Avci}, \citenamefont
  {Zeledon}, \citenamefont {Song}, \citenamefont {Gonzalez-Fuentes},
  \citenamefont {Mendil}, \citenamefont {Gambardella}, \citenamefont {Veis},
  \citenamefont {Garcia}, \citenamefont {Beach},\ and\ \citenamefont
  {Ross}}]{Rosenberg2018}%
  \BibitemOpen
  \bibfield  {author} {\bibinfo {author} {\bibfnamefont {E.~R.}\ \bibnamefont
  {Rosenberg}}, \bibinfo {author} {\bibfnamefont {L.}~\bibnamefont {Beran}},
  \bibinfo {author} {\bibfnamefont {C.~O.}\ \bibnamefont {Avci}}, \bibinfo
  {author} {\bibfnamefont {C.}~\bibnamefont {Zeledon}}, \bibinfo {author}
  {\bibfnamefont {B.}~\bibnamefont {Song}}, \bibinfo {author} {\bibfnamefont
  {C.}~\bibnamefont {Gonzalez-Fuentes}}, \bibinfo {author} {\bibfnamefont
  {J.}~\bibnamefont {Mendil}}, \bibinfo {author} {\bibfnamefont
  {P.}~\bibnamefont {Gambardella}}, \bibinfo {author} {\bibfnamefont
  {M.}~\bibnamefont {Veis}}, \bibinfo {author} {\bibfnamefont {C.}~\bibnamefont
  {Garcia}}, \bibinfo {author} {\bibfnamefont {G.~S.~D.}\ \bibnamefont
  {Beach}}, \ and\ \bibinfo {author} {\bibfnamefont {C.~A.}\ \bibnamefont
  {Ross}},\ }\bibfield  {title} {\enquote {\bibinfo {title} {Magnetism and spin
  transport in rare-earth-rich epitaxial terbium and europium iron garnet
  films},}\ }\href {\doibase 10.1103/physrevmaterials.2.094405} {\bibfield
  {journal} {\bibinfo  {journal} {Physical Review Materials}\ }\textbf
  {\bibinfo {volume} {2}},\ \bibinfo {pages} {094405} (\bibinfo {year}
  {2018})}\BibitemShut {NoStop}%
\bibitem [{\citenamefont {Ortiz}\ \emph {et~al.}(2018)\citenamefont {Ortiz},
  \citenamefont {Aldosary}, \citenamefont {Li}, \citenamefont {Xu},
  \citenamefont {Lohmann}, \citenamefont {Sellappan}, \citenamefont {Kodera},
  \citenamefont {Garay},\ and\ \citenamefont {Shi}}]{Ortiz2018}%
  \BibitemOpen
  \bibfield  {author} {\bibinfo {author} {\bibfnamefont {V.~H.}\ \bibnamefont
  {Ortiz}}, \bibinfo {author} {\bibfnamefont {M.}~\bibnamefont {Aldosary}},
  \bibinfo {author} {\bibfnamefont {J.}~\bibnamefont {Li}}, \bibinfo {author}
  {\bibfnamefont {Y.}~\bibnamefont {Xu}}, \bibinfo {author} {\bibfnamefont
  {M.~I.}\ \bibnamefont {Lohmann}}, \bibinfo {author} {\bibfnamefont
  {P.}~\bibnamefont {Sellappan}}, \bibinfo {author} {\bibfnamefont
  {Y.}~\bibnamefont {Kodera}}, \bibinfo {author} {\bibfnamefont {J.~E.}\
  \bibnamefont {Garay}}, \ and\ \bibinfo {author} {\bibfnamefont
  {J.}~\bibnamefont {Shi}},\ }\bibfield  {title} {\enquote {\bibinfo {title}
  {Systematic control of strain-induced perpendicular magnetic anisotropy in
  epitaxial europium and terbium iron garnet thin films},}\ }\href {\doibase
  10.1063/1.5078645} {\bibfield  {journal} {\bibinfo  {journal} {{APL}
  Materials}\ }\textbf {\bibinfo {volume} {6}},\ \bibinfo {pages} {121113}
  (\bibinfo {year} {2018})}\BibitemShut {NoStop}%
\bibitem [{\citenamefont {Bauer}, \citenamefont {Rosenberg},\ and\
  \citenamefont {Ross}(2019)}]{Bauer2019}%
  \BibitemOpen
  \bibfield  {author} {\bibinfo {author} {\bibfnamefont {J.~J.}\ \bibnamefont
  {Bauer}}, \bibinfo {author} {\bibfnamefont {E.~R.}\ \bibnamefont
  {Rosenberg}}, \ and\ \bibinfo {author} {\bibfnamefont {C.~A.}\ \bibnamefont
  {Ross}},\ }\bibfield  {title} {\enquote {\bibinfo {title} {Perpendicular
  magnetic anisotropy and spin mixing conductance in polycrystalline europium
  iron garnet thin films},}\ }\href@noop {} {\bibfield  {journal} {\bibinfo
  {journal} {Applied Physics Letters}\ }\textbf {\bibinfo {volume} {114}}
  (\bibinfo {year} {2019})}\BibitemShut {NoStop}%
\bibitem [{\citenamefont {Avci}\ \emph {et~al.}(2016)\citenamefont {Avci},
  \citenamefont {Quindeau}, \citenamefont {Pai}, \citenamefont {Mann},
  \citenamefont {Caretta}, \citenamefont {Tang}, \citenamefont {Onbasli},
  \citenamefont {Ross},\ and\ \citenamefont {Beach}}]{Avci2016}%
  \BibitemOpen
  \bibfield  {author} {\bibinfo {author} {\bibfnamefont {C.~O.}\ \bibnamefont
  {Avci}}, \bibinfo {author} {\bibfnamefont {A.}~\bibnamefont {Quindeau}},
  \bibinfo {author} {\bibfnamefont {C.-F.}\ \bibnamefont {Pai}}, \bibinfo
  {author} {\bibfnamefont {M.}~\bibnamefont {Mann}}, \bibinfo {author}
  {\bibfnamefont {L.}~\bibnamefont {Caretta}}, \bibinfo {author} {\bibfnamefont
  {A.~S.}\ \bibnamefont {Tang}}, \bibinfo {author} {\bibfnamefont {M.~C.}\
  \bibnamefont {Onbasli}}, \bibinfo {author} {\bibfnamefont {C.~A.}\
  \bibnamefont {Ross}}, \ and\ \bibinfo {author} {\bibfnamefont {G.~S.~D.}\
  \bibnamefont {Beach}},\ }\bibfield  {title} {\enquote {\bibinfo {title}
  {Current-induced switching in a magnetic insulator},}\ }\href {\doibase
  10.1038/nmat4812} {\bibfield  {journal} {\bibinfo  {journal} {Nature
  Materials}\ }\textbf {\bibinfo {volume} {16}},\ \bibinfo {pages} {309--314}
  (\bibinfo {year} {2016})}\BibitemShut {NoStop}%
\bibitem [{\citenamefont {Vélez}\ \emph {et~al.}(2019)\citenamefont {Vélez},
  \citenamefont {Schaab}, \citenamefont {Wörnle}, \citenamefont {Müller},
  \citenamefont {Gradauskaite}, \citenamefont {Welter}, \citenamefont
  {Gutgsell}, \citenamefont {Nistor}, \citenamefont {Degen}, \citenamefont
  {Trassin}, \citenamefont {Fiebig},\ and\ \citenamefont
  {Gambardella}}]{Velez2019}%
  \BibitemOpen
  \bibfield  {author} {\bibinfo {author} {\bibfnamefont {S.}~\bibnamefont
  {Vélez}}, \bibinfo {author} {\bibfnamefont {J.}~\bibnamefont {Schaab}},
  \bibinfo {author} {\bibfnamefont {M.~S.}\ \bibnamefont {Wörnle}}, \bibinfo
  {author} {\bibfnamefont {M.}~\bibnamefont {Müller}}, \bibinfo {author}
  {\bibfnamefont {E.}~\bibnamefont {Gradauskaite}}, \bibinfo {author}
  {\bibfnamefont {P.}~\bibnamefont {Welter}}, \bibinfo {author} {\bibfnamefont
  {C.}~\bibnamefont {Gutgsell}}, \bibinfo {author} {\bibfnamefont
  {C.}~\bibnamefont {Nistor}}, \bibinfo {author} {\bibfnamefont {C.~L.}\
  \bibnamefont {Degen}}, \bibinfo {author} {\bibfnamefont {M.}~\bibnamefont
  {Trassin}}, \bibinfo {author} {\bibfnamefont {M.}~\bibnamefont {Fiebig}}, \
  and\ \bibinfo {author} {\bibfnamefont {P.}~\bibnamefont {Gambardella}},\
  }\bibfield  {title} {\enquote {\bibinfo {title} {High-speed domain wall
  racetracks in a magnetic insulator},}\ }\href@noop {} {\bibfield  {journal}
  {\bibinfo  {journal} {Nature Communications}\ }\textbf {\bibinfo {volume}
  {10}} (\bibinfo {year} {2019})}\BibitemShut {NoStop}%
\bibitem [{\citenamefont {Avci}\ \emph {et~al.}(2019)\citenamefont {Avci},
  \citenamefont {Rosenberg}, \citenamefont {Caretta}, \citenamefont {Büttner},
  \citenamefont {Mann}, \citenamefont {Marcus}, \citenamefont {Bono},
  \citenamefont {Ross},\ and\ \citenamefont {Beach}}]{Avci2019}%
  \BibitemOpen
  \bibfield  {author} {\bibinfo {author} {\bibfnamefont {C.~O.}\ \bibnamefont
  {Avci}}, \bibinfo {author} {\bibfnamefont {E.}~\bibnamefont {Rosenberg}},
  \bibinfo {author} {\bibfnamefont {L.}~\bibnamefont {Caretta}}, \bibinfo
  {author} {\bibfnamefont {F.}~\bibnamefont {Büttner}}, \bibinfo {author}
  {\bibfnamefont {M.}~\bibnamefont {Mann}}, \bibinfo {author} {\bibfnamefont
  {C.}~\bibnamefont {Marcus}}, \bibinfo {author} {\bibfnamefont
  {D.}~\bibnamefont {Bono}}, \bibinfo {author} {\bibfnamefont {C.~A.}\
  \bibnamefont {Ross}}, \ and\ \bibinfo {author} {\bibfnamefont {G.~S.~D.}\
  \bibnamefont {Beach}},\ }\bibfield  {title} {\enquote {\bibinfo {title}
  {Interface-driven chiral magnetism and current-driven domain walls in
  insulating magnetic garnets},}\ }\href {\doibase 10.1038/s41565-019-0421-2}
  {\bibfield  {journal} {\bibinfo  {journal} {Nature Nanotechnology}\ }\textbf
  {\bibinfo {volume} {14}},\ \bibinfo {pages} {561--566} (\bibinfo {year}
  {2019})}\BibitemShut {NoStop}%
\bibitem [{\citenamefont {Klingler}\ \emph {et~al.}(2015)\citenamefont
  {Klingler}, \citenamefont {Pirro}, \citenamefont {Brächer}, \citenamefont
  {Leven}, \citenamefont {Hillebrands},\ and\ \citenamefont
  {Chumak}}]{Klingler2015}%
  \BibitemOpen
  \bibfield  {author} {\bibinfo {author} {\bibfnamefont {S.}~\bibnamefont
  {Klingler}}, \bibinfo {author} {\bibfnamefont {P.}~\bibnamefont {Pirro}},
  \bibinfo {author} {\bibfnamefont {T.}~\bibnamefont {Brächer}}, \bibinfo
  {author} {\bibfnamefont {B.}~\bibnamefont {Leven}}, \bibinfo {author}
  {\bibfnamefont {B.}~\bibnamefont {Hillebrands}}, \ and\ \bibinfo {author}
  {\bibfnamefont {A.~V.}\ \bibnamefont {Chumak}},\ }\bibfield  {title}
  {\enquote {\bibinfo {title} {Spin-wave logic devices based on isotropic
  forward volume magnetostatic waves},}\ }\href {\doibase 10.1063/1.4921850}
  {\bibfield  {journal} {\bibinfo  {journal} {Applied Physics Letters}\
  }\textbf {\bibinfo {volume} {106}} (\bibinfo {year} {2015}),\
  10.1063/1.4921850}\BibitemShut {NoStop}%
\bibitem [{\citenamefont {Shao}\ \emph {et~al.}(2019)\citenamefont {Shao},
  \citenamefont {Liu}, \citenamefont {Yu}, \citenamefont {Kim}, \citenamefont
  {Che}, \citenamefont {Tang}, \citenamefont {He}, \citenamefont {Tserkovnyak},
  \citenamefont {Shi},\ and\ \citenamefont {Wang}}]{Shao2019}%
  \BibitemOpen
  \bibfield  {author} {\bibinfo {author} {\bibfnamefont {Q.}~\bibnamefont
  {Shao}}, \bibinfo {author} {\bibfnamefont {Y.}~\bibnamefont {Liu}}, \bibinfo
  {author} {\bibfnamefont {G.}~\bibnamefont {Yu}}, \bibinfo {author}
  {\bibfnamefont {S.~K.}\ \bibnamefont {Kim}}, \bibinfo {author} {\bibfnamefont
  {X.}~\bibnamefont {Che}}, \bibinfo {author} {\bibfnamefont {C.}~\bibnamefont
  {Tang}}, \bibinfo {author} {\bibfnamefont {Q.~L.}\ \bibnamefont {He}},
  \bibinfo {author} {\bibfnamefont {Y.}~\bibnamefont {Tserkovnyak}}, \bibinfo
  {author} {\bibfnamefont {J.}~\bibnamefont {Shi}}, \ and\ \bibinfo {author}
  {\bibfnamefont {K.~L.}\ \bibnamefont {Wang}},\ }\bibfield  {title} {\enquote
  {\bibinfo {title} {Topological {H}all effect at above room temperature in
  heterostructures composed of a magnetic insulator and a heavy metal},}\
  }\href {\doibase 10.1038/s41928-019-0246-x} {\bibfield  {journal} {\bibinfo
  {journal} {Nature Electronics}\ }\textbf {\bibinfo {volume} {2}},\ \bibinfo
  {pages} {182--186} (\bibinfo {year} {2019})}\BibitemShut {NoStop}%
\bibitem [{\citenamefont {Caretta}\ \emph {et~al.}(2020)\citenamefont
  {Caretta}, \citenamefont {Rosenberg}, \citenamefont {Büttner}, \citenamefont
  {Fakhrul}, \citenamefont {Gargiani}, \citenamefont {Valvidares},
  \citenamefont {Chen}, \citenamefont {Reddy}, \citenamefont {Muller},
  \citenamefont {Ross},\ and\ \citenamefont {Beach}}]{Caretta2020}%
  \BibitemOpen
  \bibfield  {author} {\bibinfo {author} {\bibfnamefont {L.}~\bibnamefont
  {Caretta}}, \bibinfo {author} {\bibfnamefont {E.}~\bibnamefont {Rosenberg}},
  \bibinfo {author} {\bibfnamefont {F.}~\bibnamefont {Büttner}}, \bibinfo
  {author} {\bibfnamefont {T.}~\bibnamefont {Fakhrul}}, \bibinfo {author}
  {\bibfnamefont {P.}~\bibnamefont {Gargiani}}, \bibinfo {author}
  {\bibfnamefont {M.}~\bibnamefont {Valvidares}}, \bibinfo {author}
  {\bibfnamefont {Z.}~\bibnamefont {Chen}}, \bibinfo {author} {\bibfnamefont
  {P.}~\bibnamefont {Reddy}}, \bibinfo {author} {\bibfnamefont {D.~A.}\
  \bibnamefont {Muller}}, \bibinfo {author} {\bibfnamefont {C.~A.}\
  \bibnamefont {Ross}}, \ and\ \bibinfo {author} {\bibfnamefont {G.~S.~D.}\
  \bibnamefont {Beach}},\ }\bibfield  {title} {\enquote {\bibinfo {title}
  {Interfacial {D}zyaloshinskii-{M}oriya interaction arising from rare-earth
  orbital magnetism in insulating magnetic oxides},}\ }\href {\doibase
  10.1038/s41467-020-14924-7} {\bibfield  {journal} {\bibinfo  {journal}
  {Nature Communications}\ }\textbf {\bibinfo {volume} {11}},\ \bibinfo {pages}
  {1090} (\bibinfo {year} {2020})}\BibitemShut {NoStop}%
\bibitem [{\citenamefont {Wang}\ \emph {et~al.}(2020)\citenamefont {Wang},
  \citenamefont {Chen}, \citenamefont {Liu}, \citenamefont {Zhang},
  \citenamefont {Baumgaertl}, \citenamefont {Guo}, \citenamefont {Li},
  \citenamefont {Liu}, \citenamefont {Che}, \citenamefont {Tu}, \citenamefont
  {Liu}, \citenamefont {Gao}, \citenamefont {Han}, \citenamefont {Yu},
  \citenamefont {Wu}, \citenamefont {Grundler},\ and\ \citenamefont
  {Yu}}]{Wang2020b}%
  \BibitemOpen
  \bibfield  {author} {\bibinfo {author} {\bibfnamefont {H.}~\bibnamefont
  {Wang}}, \bibinfo {author} {\bibfnamefont {J.}~\bibnamefont {Chen}}, \bibinfo
  {author} {\bibfnamefont {T.}~\bibnamefont {Liu}}, \bibinfo {author}
  {\bibfnamefont {J.}~\bibnamefont {Zhang}}, \bibinfo {author} {\bibfnamefont
  {K.}~\bibnamefont {Baumgaertl}}, \bibinfo {author} {\bibfnamefont
  {C.}~\bibnamefont {Guo}}, \bibinfo {author} {\bibfnamefont {Y.}~\bibnamefont
  {Li}}, \bibinfo {author} {\bibfnamefont {C.}~\bibnamefont {Liu}}, \bibinfo
  {author} {\bibfnamefont {P.}~\bibnamefont {Che}}, \bibinfo {author}
  {\bibfnamefont {S.}~\bibnamefont {Tu}}, \bibinfo {author} {\bibfnamefont
  {S.}~\bibnamefont {Liu}}, \bibinfo {author} {\bibfnamefont {P.}~\bibnamefont
  {Gao}}, \bibinfo {author} {\bibfnamefont {X.}~\bibnamefont {Han}}, \bibinfo
  {author} {\bibfnamefont {D.}~\bibnamefont {Yu}}, \bibinfo {author}
  {\bibfnamefont {M.}~\bibnamefont {Wu}}, \bibinfo {author} {\bibfnamefont
  {D.}~\bibnamefont {Grundler}}, \ and\ \bibinfo {author} {\bibfnamefont
  {H.}~\bibnamefont {Yu}},\ }\bibfield  {title} {\enquote {\bibinfo {title}
  {Chiral spin-wave velocities induced by all-garnet interfacial
  {D}zyaloshinskii-{M}oriya interaction in ultrathin yttrium iron garnet
  films},}\ }\href {\doibase 10.1103/physrevlett.124.027203} {\bibfield
  {journal} {\bibinfo  {journal} {Physical Review Letters}\ }\textbf {\bibinfo
  {volume} {124}},\ \bibinfo {pages} {027203} (\bibinfo {year}
  {2020})}\BibitemShut {NoStop}%
\bibitem [{\citenamefont {Schlitz}\ \emph {et~al.}(2021)\citenamefont
  {Schlitz}, \citenamefont {V{\'{e}}lez}, \citenamefont {Kamra}, \citenamefont
  {Lambert}, \citenamefont {Lammel}, \citenamefont {Goennenwein},\ and\
  \citenamefont {Gambardella}}]{Schlitz2021}%
  \BibitemOpen
  \bibfield  {author} {\bibinfo {author} {\bibfnamefont {R.}~\bibnamefont
  {Schlitz}}, \bibinfo {author} {\bibfnamefont {S.}~\bibnamefont
  {V{\'{e}}lez}}, \bibinfo {author} {\bibfnamefont {A.}~\bibnamefont {Kamra}},
  \bibinfo {author} {\bibfnamefont {C.-H.}\ \bibnamefont {Lambert}}, \bibinfo
  {author} {\bibfnamefont {M.}~\bibnamefont {Lammel}}, \bibinfo {author}
  {\bibfnamefont {S.~T.}\ \bibnamefont {Goennenwein}}, \ and\ \bibinfo {author}
  {\bibfnamefont {P.}~\bibnamefont {Gambardella}},\ }\bibfield  {title}
  {\enquote {\bibinfo {title} {Control of nonlocal magnon spin transport via
  magnon drift currents},}\ }\href {\doibase 10.1103/physrevlett.126.257201}
  {\bibfield  {journal} {\bibinfo  {journal} {Physical Review Letters}\
  }\textbf {\bibinfo {volume} {126}},\ \bibinfo {pages} {257201} (\bibinfo
  {year} {2021})}\BibitemShut {NoStop}%
\bibitem [{\citenamefont {Ding}\ \emph {et~al.}(2019)\citenamefont {Ding},
  \citenamefont {Ross}, \citenamefont {Lebrun}, \citenamefont {Becker},
  \citenamefont {Lee}, \citenamefont {Boventer}, \citenamefont {Das},
  \citenamefont {Kurokawa}, \citenamefont {Gupta}, \citenamefont {Yang},
  \citenamefont {Jakob},\ and\ \citenamefont {Kläui}}]{Ding2019}%
  \BibitemOpen
  \bibfield  {author} {\bibinfo {author} {\bibfnamefont {S.}~\bibnamefont
  {Ding}}, \bibinfo {author} {\bibfnamefont {A.}~\bibnamefont {Ross}}, \bibinfo
  {author} {\bibfnamefont {R.}~\bibnamefont {Lebrun}}, \bibinfo {author}
  {\bibfnamefont {S.}~\bibnamefont {Becker}}, \bibinfo {author} {\bibfnamefont
  {K.}~\bibnamefont {Lee}}, \bibinfo {author} {\bibfnamefont {I.}~\bibnamefont
  {Boventer}}, \bibinfo {author} {\bibfnamefont {S.}~\bibnamefont {Das}},
  \bibinfo {author} {\bibfnamefont {Y.}~\bibnamefont {Kurokawa}}, \bibinfo
  {author} {\bibfnamefont {S.}~\bibnamefont {Gupta}}, \bibinfo {author}
  {\bibfnamefont {J.}~\bibnamefont {Yang}}, \bibinfo {author} {\bibfnamefont
  {G.}~\bibnamefont {Jakob}}, \ and\ \bibinfo {author} {\bibfnamefont
  {M.}~\bibnamefont {Kläui}},\ }\bibfield  {title} {\enquote {\bibinfo {title}
  {Interfacial {D}zyaloshinskii-{M}oriya interaction and chiral magnetic
  textures in a ferrimagnetic insulator},}\ }\href {\doibase
  10.1103/physrevb.100.100406} {\bibfield  {journal} {\bibinfo  {journal}
  {Physical Review B}\ }\textbf {\bibinfo {volume} {100}},\ \bibinfo {pages}
  {100406} (\bibinfo {year} {2019})}\BibitemShut {NoStop}%
\bibitem [{\citenamefont {Büttner}\ \emph {et~al.}(2020)\citenamefont
  {Büttner}, \citenamefont {Mawass}, \citenamefont {Bauer}, \citenamefont
  {Rosenberg}, \citenamefont {Caretta}, \citenamefont {Avci}, \citenamefont
  {Gräfe}, \citenamefont {Finizio}, \citenamefont {Vaz}, \citenamefont
  {Novakovic}, \citenamefont {Weigand}, \citenamefont {Litzius}, \citenamefont
  {Förster}, \citenamefont {Träger}, \citenamefont {Gro{\ss}}, \citenamefont
  {Suzuki}, \citenamefont {Huang}, \citenamefont {Bartell}, \citenamefont
  {Kronast}, \citenamefont {Raabe}, \citenamefont {Schütz}, \citenamefont
  {Ross},\ and\ \citenamefont {Beach}}]{Buttner2020}%
  \BibitemOpen
  \bibfield  {author} {\bibinfo {author} {\bibfnamefont {F.}~\bibnamefont
  {Büttner}}, \bibinfo {author} {\bibfnamefont {M.~A.}\ \bibnamefont
  {Mawass}}, \bibinfo {author} {\bibfnamefont {J.}~\bibnamefont {Bauer}},
  \bibinfo {author} {\bibfnamefont {E.}~\bibnamefont {Rosenberg}}, \bibinfo
  {author} {\bibfnamefont {L.}~\bibnamefont {Caretta}}, \bibinfo {author}
  {\bibfnamefont {C.~O.}\ \bibnamefont {Avci}}, \bibinfo {author}
  {\bibfnamefont {J.}~\bibnamefont {Gräfe}}, \bibinfo {author} {\bibfnamefont
  {S.}~\bibnamefont {Finizio}}, \bibinfo {author} {\bibfnamefont {C.~A.~F.}\
  \bibnamefont {Vaz}}, \bibinfo {author} {\bibfnamefont {N.}~\bibnamefont
  {Novakovic}}, \bibinfo {author} {\bibfnamefont {M.}~\bibnamefont {Weigand}},
  \bibinfo {author} {\bibfnamefont {K.}~\bibnamefont {Litzius}}, \bibinfo
  {author} {\bibfnamefont {J.}~\bibnamefont {Förster}}, \bibinfo {author}
  {\bibfnamefont {N.}~\bibnamefont {Träger}}, \bibinfo {author} {\bibfnamefont
  {F.}~\bibnamefont {Gro{\ss}}}, \bibinfo {author} {\bibfnamefont
  {D.}~\bibnamefont {Suzuki}}, \bibinfo {author} {\bibfnamefont
  {M.}~\bibnamefont {Huang}}, \bibinfo {author} {\bibfnamefont
  {J.}~\bibnamefont {Bartell}}, \bibinfo {author} {\bibfnamefont
  {F.}~\bibnamefont {Kronast}}, \bibinfo {author} {\bibfnamefont
  {J.}~\bibnamefont {Raabe}}, \bibinfo {author} {\bibfnamefont
  {G.}~\bibnamefont {Schütz}}, \bibinfo {author} {\bibfnamefont {C.~A.}\
  \bibnamefont {Ross}}, \ and\ \bibinfo {author} {\bibfnamefont {G.~S.~D.}\
  \bibnamefont {Beach}},\ }\bibfield  {title} {\enquote {\bibinfo {title}
  {Thermal nucleation and high-resolution imaging of submicrometer magnetic
  bubbles in thin thulium iron garnet films with perpendicular anisotropy},}\
  }\href@noop {} {\bibfield  {journal} {\bibinfo  {journal} {Physical Review
  Materials}\ }\textbf {\bibinfo {volume} {4}} (\bibinfo {year}
  {2020})}\BibitemShut {NoStop}%
\bibitem [{\citenamefont {V{\'{e}}lez}\ \emph {et~al.}(2022)\citenamefont
  {V{\'{e}}lez}, \citenamefont {Ruiz-G{\'{o}}mez}, \citenamefont {Schaab},
  \citenamefont {Gradauskaite}, \citenamefont {Wörnle}, \citenamefont
  {Welter}, \citenamefont {Jacot}, \citenamefont {Degen}, \citenamefont
  {Trassin}, \citenamefont {Fiebig},\ and\ \citenamefont
  {Gambardella}}]{Velez2022}%
  \BibitemOpen
  \bibfield  {author} {\bibinfo {author} {\bibfnamefont {S.}~\bibnamefont
  {V{\'{e}}lez}}, \bibinfo {author} {\bibfnamefont {S.}~\bibnamefont
  {Ruiz-G{\'{o}}mez}}, \bibinfo {author} {\bibfnamefont {J.}~\bibnamefont
  {Schaab}}, \bibinfo {author} {\bibfnamefont {E.}~\bibnamefont
  {Gradauskaite}}, \bibinfo {author} {\bibfnamefont {M.~S.}\ \bibnamefont
  {Wörnle}}, \bibinfo {author} {\bibfnamefont {P.}~\bibnamefont {Welter}},
  \bibinfo {author} {\bibfnamefont {B.~J.}\ \bibnamefont {Jacot}}, \bibinfo
  {author} {\bibfnamefont {C.~L.}\ \bibnamefont {Degen}}, \bibinfo {author}
  {\bibfnamefont {M.}~\bibnamefont {Trassin}}, \bibinfo {author} {\bibfnamefont
  {M.}~\bibnamefont {Fiebig}}, \ and\ \bibinfo {author} {\bibfnamefont
  {P.}~\bibnamefont {Gambardella}},\ }\bibfield  {title} {\enquote {\bibinfo
  {title} {Current-driven dynamics and ratchet effect of skyrmion bubbles in a
  ferrimagnetic insulator},}\ }\href {\doibase 10.1038/s41565-022-01144-x}
  {\bibfield  {journal} {\bibinfo  {journal} {Nature Nanotechnology}\ }\textbf
  {\bibinfo {volume} {17}},\ \bibinfo {pages} {834--841} (\bibinfo {year}
  {2022})}\BibitemShut {NoStop}%
\bibitem [{\citenamefont {Fakhrul}\ \emph {et~al.}(2024)\citenamefont
  {Fakhrul}, \citenamefont {Khurana}, \citenamefont {Lee}, \citenamefont
  {Huang}, \citenamefont {Nembach}, \citenamefont {Beach},\ and\ \citenamefont
  {Ross}}]{Fakhrul2024}%
  \BibitemOpen
  \bibfield  {author} {\bibinfo {author} {\bibfnamefont {T.}~\bibnamefont
  {Fakhrul}}, \bibinfo {author} {\bibfnamefont {B.}~\bibnamefont {Khurana}},
  \bibinfo {author} {\bibfnamefont {B.~H.}\ \bibnamefont {Lee}}, \bibinfo
  {author} {\bibfnamefont {S.}~\bibnamefont {Huang}}, \bibinfo {author}
  {\bibfnamefont {H.~T.}\ \bibnamefont {Nembach}}, \bibinfo {author}
  {\bibfnamefont {G.~S.~D.}\ \bibnamefont {Beach}}, \ and\ \bibinfo {author}
  {\bibfnamefont {C.~A.}\ \bibnamefont {Ross}},\ }\bibfield  {title} {\enquote
  {\bibinfo {title} {Damping and interfacial {D}zyaloshinskii–{M}oriya
  interaction in thulium iron garnet/bismuth-substituted yttrium iron garnet
  bilayers},}\ }\href {\doibase 10.1021/acsami.3c14706} {\bibfield  {journal}
  {\bibinfo  {journal} {ACS Applied Materials \& Interfaces}\ }\textbf
  {\bibinfo {volume} {16}},\ \bibinfo {pages} {2489--2496} (\bibinfo {year}
  {2024})}\BibitemShut {NoStop}%
\bibitem [{\citenamefont {Sakamaki}\ \emph {et~al.}(2012)\citenamefont
  {Sakamaki}, \citenamefont {Amemiya}, \citenamefont {Liedke}, \citenamefont
  {Fassbender}, \citenamefont {Mazalski}, \citenamefont {Sveklo},\ and\
  \citenamefont {Maziewski}}]{Sakamaki2012}%
  \BibitemOpen
  \bibfield  {author} {\bibinfo {author} {\bibfnamefont {M.}~\bibnamefont
  {Sakamaki}}, \bibinfo {author} {\bibfnamefont {K.}~\bibnamefont {Amemiya}},
  \bibinfo {author} {\bibfnamefont {M.~O.}\ \bibnamefont {Liedke}}, \bibinfo
  {author} {\bibfnamefont {J.}~\bibnamefont {Fassbender}}, \bibinfo {author}
  {\bibfnamefont {P.}~\bibnamefont {Mazalski}}, \bibinfo {author}
  {\bibfnamefont {I.}~\bibnamefont {Sveklo}}, \ and\ \bibinfo {author}
  {\bibfnamefont {A.}~\bibnamefont {Maziewski}},\ }\bibfield  {title} {\enquote
  {\bibinfo {title} {Perpendicular magnetic anisotropy in a {P}t/{C}o/{P}t
  ultrathin film arising from a lattice distortion induced by ion
  irradiation},}\ }\href {\doibase 10.1103/physrevb.86.024418} {\bibfield
  {journal} {\bibinfo  {journal} {Physical Review B}\ }\textbf {\bibinfo
  {volume} {86}},\ \bibinfo {pages} {024418} (\bibinfo {year}
  {2012})}\BibitemShut {NoStop}%
\bibitem [{\citenamefont {Balan}\ \emph {et~al.}(2023)\citenamefont {Balan},
  \citenamefont {van~der Jagt}, \citenamefont {Fassatoui}, \citenamefont
  {Garcia}, \citenamefont {Jeudy}, \citenamefont {Thiaville}, \citenamefont
  {Bonfim}, \citenamefont {Vogel}, \citenamefont {Ranno}, \citenamefont
  {Ravelosona},\ and\ \citenamefont {Pizzini}}]{Balan2023}%
  \BibitemOpen
  \bibfield  {author} {\bibinfo {author} {\bibfnamefont {C.}~\bibnamefont
  {Balan}}, \bibinfo {author} {\bibfnamefont {J.~W.}\ \bibnamefont {van~der
  Jagt}}, \bibinfo {author} {\bibfnamefont {A.}~\bibnamefont {Fassatoui}},
  \bibinfo {author} {\bibfnamefont {J.~P.}\ \bibnamefont {Garcia}}, \bibinfo
  {author} {\bibfnamefont {V.}~\bibnamefont {Jeudy}}, \bibinfo {author}
  {\bibfnamefont {A.}~\bibnamefont {Thiaville}}, \bibinfo {author}
  {\bibfnamefont {M.}~\bibnamefont {Bonfim}}, \bibinfo {author} {\bibfnamefont
  {J.}~\bibnamefont {Vogel}}, \bibinfo {author} {\bibfnamefont
  {L.}~\bibnamefont {Ranno}}, \bibinfo {author} {\bibfnamefont
  {D.}~\bibnamefont {Ravelosona}}, \ and\ \bibinfo {author} {\bibfnamefont
  {S.}~\bibnamefont {Pizzini}},\ }\bibfield  {title} {\enquote {\bibinfo
  {title} {Improving {N\'{e}}el domain walls dynamics and skyrmion stability
  using {H}e ion irradiation},}\ }\href@noop {} {\bibfield  {journal} {\bibinfo
   {journal} {Small}\ }\textbf {\bibinfo {volume} {19}} (\bibinfo {year}
  {2023})}\BibitemShut {NoStop}%
\bibitem [{\citenamefont {Diez}\ \emph {et~al.}(2019)\citenamefont {Diez},
  \citenamefont {Voto}, \citenamefont {Casiraghi}, \citenamefont {Belmeguenai},
  \citenamefont {Roussign\'e}, \citenamefont {Durin}, \citenamefont {Lamperti},
  \citenamefont {Mantovan}, \citenamefont {Sluka}, \citenamefont {Jeudy},
  \citenamefont {Liu}, \citenamefont {Stashkevich}, \citenamefont {Ch\'erif},
  \citenamefont {Langer}, \citenamefont {Ocker}, \citenamefont {Lopez-Diaz},\
  and\ \citenamefont {Ravelosona}}]{Diez2019}%
  \BibitemOpen
  \bibfield  {author} {\bibinfo {author} {\bibfnamefont {L.~H.}\ \bibnamefont
  {Diez}}, \bibinfo {author} {\bibfnamefont {M.}~\bibnamefont {Voto}}, \bibinfo
  {author} {\bibfnamefont {A.}~\bibnamefont {Casiraghi}}, \bibinfo {author}
  {\bibfnamefont {M.}~\bibnamefont {Belmeguenai}}, \bibinfo {author}
  {\bibfnamefont {Y.}~\bibnamefont {Roussign\'e}}, \bibinfo {author}
  {\bibfnamefont {G.}~\bibnamefont {Durin}}, \bibinfo {author} {\bibfnamefont
  {A.}~\bibnamefont {Lamperti}}, \bibinfo {author} {\bibfnamefont
  {R.}~\bibnamefont {Mantovan}}, \bibinfo {author} {\bibfnamefont
  {V.}~\bibnamefont {Sluka}}, \bibinfo {author} {\bibfnamefont
  {V.}~\bibnamefont {Jeudy}}, \bibinfo {author} {\bibfnamefont {Y.~T.}\
  \bibnamefont {Liu}}, \bibinfo {author} {\bibfnamefont {A.}~\bibnamefont
  {Stashkevich}}, \bibinfo {author} {\bibfnamefont {S.~M.}\ \bibnamefont
  {Ch\'erif}}, \bibinfo {author} {\bibfnamefont {J.}~\bibnamefont {Langer}},
  \bibinfo {author} {\bibfnamefont {B.}~\bibnamefont {Ocker}}, \bibinfo
  {author} {\bibfnamefont {L.}~\bibnamefont {Lopez-Diaz}}, \ and\ \bibinfo
  {author} {\bibfnamefont {D.}~\bibnamefont {Ravelosona}},\ }\bibfield  {title}
  {\enquote {\bibinfo {title} {Enhancement of the {D}zyaloshinskii-{M}oriya
  interaction and domain wall velocity through interface intermixing in
  {Ta/CoFeB/MgO}},}\ }\href@noop {} {\bibfield  {journal} {\bibinfo  {journal}
  {Phys. Rev. B}\ }\textbf {\bibinfo {volume} {99}},\ \bibinfo {pages} {054431}
  (\bibinfo {year} {2019})}\BibitemShut {NoStop}%
\bibitem [{\citenamefont {Juge}\ \emph {et~al.}(2021)\citenamefont {Juge},
  \citenamefont {Bairagi}, \citenamefont {Rana}, \citenamefont {Vogel},
  \citenamefont {Sall}, \citenamefont {Mailly}, \citenamefont {Pham},
  \citenamefont {Zhang}, \citenamefont {Sisodia}, \citenamefont {Foerster},
  \citenamefont {Aballe}, \citenamefont {Belmeguenai}, \citenamefont
  {Roussign{\'{e}}}, \citenamefont {Auffret}, \citenamefont {Buda-Prejbeanu},
  \citenamefont {Gaudin}, \citenamefont {Ravelosona},\ and\ \citenamefont
  {Boulle}}]{Juge2021}%
  \BibitemOpen
  \bibfield  {author} {\bibinfo {author} {\bibfnamefont {R.}~\bibnamefont
  {Juge}}, \bibinfo {author} {\bibfnamefont {K.}~\bibnamefont {Bairagi}},
  \bibinfo {author} {\bibfnamefont {K.~G.}\ \bibnamefont {Rana}}, \bibinfo
  {author} {\bibfnamefont {J.}~\bibnamefont {Vogel}}, \bibinfo {author}
  {\bibfnamefont {M.}~\bibnamefont {Sall}}, \bibinfo {author} {\bibfnamefont
  {D.}~\bibnamefont {Mailly}}, \bibinfo {author} {\bibfnamefont {V.~T.}\
  \bibnamefont {Pham}}, \bibinfo {author} {\bibfnamefont {Q.}~\bibnamefont
  {Zhang}}, \bibinfo {author} {\bibfnamefont {N.}~\bibnamefont {Sisodia}},
  \bibinfo {author} {\bibfnamefont {M.}~\bibnamefont {Foerster}}, \bibinfo
  {author} {\bibfnamefont {L.}~\bibnamefont {Aballe}}, \bibinfo {author}
  {\bibfnamefont {M.}~\bibnamefont {Belmeguenai}}, \bibinfo {author}
  {\bibfnamefont {Y.}~\bibnamefont {Roussign{\'{e}}}}, \bibinfo {author}
  {\bibfnamefont {S.}~\bibnamefont {Auffret}}, \bibinfo {author} {\bibfnamefont
  {L.~D.}\ \bibnamefont {Buda-Prejbeanu}}, \bibinfo {author} {\bibfnamefont
  {G.}~\bibnamefont {Gaudin}}, \bibinfo {author} {\bibfnamefont
  {D.}~\bibnamefont {Ravelosona}}, \ and\ \bibinfo {author} {\bibfnamefont
  {O.}~\bibnamefont {Boulle}},\ }\bibfield  {title} {\enquote {\bibinfo {title}
  {Helium ions put magnetic skyrmions on the track},}\ }\href@noop {}
  {\bibfield  {journal} {\bibinfo  {journal} {Nano Letters}\ } (\bibinfo {year}
  {2021})}\BibitemShut {NoStop}%
\bibitem [{\citenamefont {Gieniusz}\ \emph {et~al.}(2021)\citenamefont
  {Gieniusz}, \citenamefont {Mazalski}, \citenamefont {Guzowska}, \citenamefont
  {Sveklo}, \citenamefont {Fassbender}, \citenamefont {Wawro},\ and\
  \citenamefont {Maziewski}}]{Gieniusz2021}%
  \BibitemOpen
  \bibfield  {author} {\bibinfo {author} {\bibfnamefont {R.}~\bibnamefont
  {Gieniusz}}, \bibinfo {author} {\bibfnamefont {P.}~\bibnamefont {Mazalski}},
  \bibinfo {author} {\bibfnamefont {U.}~\bibnamefont {Guzowska}}, \bibinfo
  {author} {\bibfnamefont {I.}~\bibnamefont {Sveklo}}, \bibinfo {author}
  {\bibfnamefont {J.}~\bibnamefont {Fassbender}}, \bibinfo {author}
  {\bibfnamefont {A.}~\bibnamefont {Wawro}}, \ and\ \bibinfo {author}
  {\bibfnamefont {A.}~\bibnamefont {Maziewski}},\ }\bibfield  {title} {\enquote
  {\bibinfo {title} {Dzyaloshinskii-{M}oriya interaction and magnetic
  anisotropy in {P}t/{C}o/{A}u trilayers modified by {G}a$^+$ ion
  irradiation},}\ }\href {\doibase 10.1016/j.jmmm.2021.168160} {\bibfield
  {journal} {\bibinfo  {journal} {Journal of Magnetism and Magnetic Materials}\
  }\textbf {\bibinfo {volume} {537}},\ \bibinfo {pages} {168160} (\bibinfo
  {year} {2021})}\BibitemShut {NoStop}%
\bibitem [{\citenamefont {Nembach}\ \emph {et~al.}(2022)\citenamefont
  {Nembach}, \citenamefont {Ju{\'{e}}}, \citenamefont {Poetzger}, \citenamefont
  {Fassbender}, \citenamefont {Silva},\ and\ \citenamefont
  {Shaw}}]{Nembach2022}%
  \BibitemOpen
  \bibfield  {author} {\bibinfo {author} {\bibfnamefont {H.~T.}\ \bibnamefont
  {Nembach}}, \bibinfo {author} {\bibfnamefont {E.}~\bibnamefont {Ju{\'{e}}}},
  \bibinfo {author} {\bibfnamefont {K.}~\bibnamefont {Poetzger}}, \bibinfo
  {author} {\bibfnamefont {J.}~\bibnamefont {Fassbender}}, \bibinfo {author}
  {\bibfnamefont {T.~J.}\ \bibnamefont {Silva}}, \ and\ \bibinfo {author}
  {\bibfnamefont {J.~M.}\ \bibnamefont {Shaw}},\ }\bibfield  {title} {\enquote
  {\bibinfo {title} {Tuning of the {D}zyaloshinskii-{M}oriya interaction by
  {H}e$^+$ ion irradiation},}\ }\href@noop {} {\bibfield  {journal} {\bibinfo
  {journal} {Journal of Applied Physics}\ }\textbf {\bibinfo {volume} {131}}
  (\bibinfo {year} {2022})}\BibitemShut {NoStop}%
\bibitem [{\citenamefont {Mewes}\ \emph {et~al.}(2000)\citenamefont {Mewes},
  \citenamefont {Lopusnik}, \citenamefont {Fassbender}, \citenamefont
  {Hillebrands}, \citenamefont {Jung}, \citenamefont {Engel}, \citenamefont
  {Ehresmann},\ and\ \citenamefont {Schmoranzer}}]{Mewes2000}%
  \BibitemOpen
  \bibfield  {author} {\bibinfo {author} {\bibfnamefont {T.}~\bibnamefont
  {Mewes}}, \bibinfo {author} {\bibfnamefont {R.}~\bibnamefont {Lopusnik}},
  \bibinfo {author} {\bibfnamefont {J.}~\bibnamefont {Fassbender}}, \bibinfo
  {author} {\bibfnamefont {B.}~\bibnamefont {Hillebrands}}, \bibinfo {author}
  {\bibfnamefont {M.}~\bibnamefont {Jung}}, \bibinfo {author} {\bibfnamefont
  {D.}~\bibnamefont {Engel}}, \bibinfo {author} {\bibfnamefont
  {A.}~\bibnamefont {Ehresmann}}, \ and\ \bibinfo {author} {\bibfnamefont
  {H.}~\bibnamefont {Schmoranzer}},\ }\bibfield  {title} {\enquote {\bibinfo
  {title} {Suppression of exchange bias by ion irradiation},}\ }\href@noop {}
  {\bibfield  {journal} {\bibinfo  {journal} {Applied Physics Letters}\
  }\textbf {\bibinfo {volume} {76}},\ \bibinfo {pages} {1057--1059} (\bibinfo
  {year} {2000})}\BibitemShut {NoStop}%
\bibitem [{\citenamefont {Teixeira}\ \emph {et~al.}(2020)\citenamefont
  {Teixeira}, \citenamefont {Timopheev}, \citenamefont {Ca{\c{c}}oilo},
  \citenamefont {Cuchet}, \citenamefont {Mondaud}, \citenamefont {Childress},
  \citenamefont {Magalh{\~{a}}es}, \citenamefont {Alves},\ and\ \citenamefont
  {Sobolev}}]{Teixeira2020}%
  \BibitemOpen
  \bibfield  {author} {\bibinfo {author} {\bibfnamefont {B.~M.~S.}\
  \bibnamefont {Teixeira}}, \bibinfo {author} {\bibfnamefont {A.~A.}\
  \bibnamefont {Timopheev}}, \bibinfo {author} {\bibfnamefont {N.}~\bibnamefont
  {Ca{\c{c}}oilo}}, \bibinfo {author} {\bibfnamefont {L.}~\bibnamefont
  {Cuchet}}, \bibinfo {author} {\bibfnamefont {J.}~\bibnamefont {Mondaud}},
  \bibinfo {author} {\bibfnamefont {J.~R.}\ \bibnamefont {Childress}}, \bibinfo
  {author} {\bibfnamefont {S.}~\bibnamefont {Magalh{\~{a}}es}}, \bibinfo
  {author} {\bibfnamefont {E.}~\bibnamefont {Alves}}, \ and\ \bibinfo {author}
  {\bibfnamefont {N.~A.}\ \bibnamefont {Sobolev}},\ }\bibfield  {title}
  {\enquote {\bibinfo {title} {Ar$^+$ ion irradiation of magnetic tunnel
  junction multilayers: impact on the magnetic and electrical properties},}\
  }\href {\doibase 10.1088/1361-6463/aba38c} {\bibfield  {journal} {\bibinfo
  {journal} {Journal of Physics D: Applied Physics}\ }\textbf {\bibinfo
  {volume} {53}},\ \bibinfo {pages} {455003} (\bibinfo {year}
  {2020})}\BibitemShut {NoStop}%
\bibitem [{\citenamefont {Suzuki}(1986)}]{Suzuki1986}%
  \BibitemOpen
  \bibfield  {author} {\bibinfo {author} {\bibfnamefont {R.}~\bibnamefont
  {Suzuki}},\ }\bibfield  {title} {\enquote {\bibinfo {title} {Recent
  development in magnetic-bubble memory},}\ }\href {\doibase
  10.1109/proc.1986.13670} {\bibfield  {journal} {\bibinfo  {journal}
  {Proceedings of the {IEEE}}\ }\textbf {\bibinfo {volume} {74}},\ \bibinfo
  {pages} {1582--1590} (\bibinfo {year} {1986})}\BibitemShut {NoStop}%
\bibitem [{\citenamefont {Guzman}\ \emph {et~al.}(1983)\citenamefont {Guzman},
  \citenamefont {Krafft}, \citenamefont {Wang},\ and\ \citenamefont
  {Kryder}}]{Guzman1983}%
  \BibitemOpen
  \bibfield  {author} {\bibinfo {author} {\bibfnamefont {A.~M.}\ \bibnamefont
  {Guzman}}, \bibinfo {author} {\bibfnamefont {C.~S.}\ \bibnamefont {Krafft}},
  \bibinfo {author} {\bibfnamefont {X.}~\bibnamefont {Wang}}, \ and\ \bibinfo
  {author} {\bibfnamefont {M.~H.}\ \bibnamefont {Kryder}},\ }\bibfield  {title}
  {\enquote {\bibinfo {title} {The effect of ion implantation on epitaxial
  magnetic garnet thin films},}\ }\href {\doibase 10.1016/0167-5087(83)90929-8}
  {\bibfield  {journal} {\bibinfo  {journal} {Nuclear Instruments and Methods
  in Physics Research}\ }\textbf {\bibinfo {volume} {209-210}},\ \bibinfo
  {pages} {1121--1127} (\bibinfo {year} {1983})}\BibitemShut {NoStop}%
\bibitem [{\citenamefont {Ruane}\ \emph {et~al.}(2017)\citenamefont {Ruane},
  \citenamefont {White}, \citenamefont {Brangham}, \citenamefont {Meng},
  \citenamefont {Pelekhov}, \citenamefont {Yang},\ and\ \citenamefont
  {Hammel}}]{Ruane2017}%
  \BibitemOpen
  \bibfield  {author} {\bibinfo {author} {\bibfnamefont {W.~T.}\ \bibnamefont
  {Ruane}}, \bibinfo {author} {\bibfnamefont {S.~P.}\ \bibnamefont {White}},
  \bibinfo {author} {\bibfnamefont {J.~T.}\ \bibnamefont {Brangham}}, \bibinfo
  {author} {\bibfnamefont {K.~Y.}\ \bibnamefont {Meng}}, \bibinfo {author}
  {\bibfnamefont {D.~V.}\ \bibnamefont {Pelekhov}}, \bibinfo {author}
  {\bibfnamefont {F.~Y.}\ \bibnamefont {Yang}}, \ and\ \bibinfo {author}
  {\bibfnamefont {P.~C.}\ \bibnamefont {Hammel}},\ }\bibfield  {title}
  {\enquote {\bibinfo {title} {Controlling and patterning the effective
  magnetization in $\mathrm{{Y}_{3}{Fe}_{5}{O}_{12}}$ thin films using ion
  irradiation},}\ }\href@noop {} {\bibfield  {journal} {\bibinfo  {journal}
  {{AIP} Advances}\ }\textbf {\bibinfo {volume} {8}} (\bibinfo {year}
  {2017})}\BibitemShut {NoStop}%
\bibitem [{\citenamefont {Kiechle}\ \emph {et~al.}(2023)\citenamefont
  {Kiechle}, \citenamefont {Papp}, \citenamefont {Mendisch}, \citenamefont
  {Ahrens}, \citenamefont {Golibrzuch}, \citenamefont {Bernstein},
  \citenamefont {Porod}, \citenamefont {Csaba},\ and\ \citenamefont
  {Becherer}}]{Kiechle2023}%
  \BibitemOpen
  \bibfield  {author} {\bibinfo {author} {\bibfnamefont {M.}~\bibnamefont
  {Kiechle}}, \bibinfo {author} {\bibfnamefont {A.}~\bibnamefont {Papp}},
  \bibinfo {author} {\bibfnamefont {S.}~\bibnamefont {Mendisch}}, \bibinfo
  {author} {\bibfnamefont {V.}~\bibnamefont {Ahrens}}, \bibinfo {author}
  {\bibfnamefont {M.}~\bibnamefont {Golibrzuch}}, \bibinfo {author}
  {\bibfnamefont {G.~H.}\ \bibnamefont {Bernstein}}, \bibinfo {author}
  {\bibfnamefont {W.}~\bibnamefont {Porod}}, \bibinfo {author} {\bibfnamefont
  {G.}~\bibnamefont {Csaba}}, \ and\ \bibinfo {author} {\bibfnamefont
  {M.}~\bibnamefont {Becherer}},\ }\bibfield  {title} {\enquote {\bibinfo
  {title} {Spin-wave optics in {YIG} realized by ion-beam irradiation},}\
  }\href@noop {} {\bibfield  {journal} {\bibinfo  {journal} {Small}\ }\textbf
  {\bibinfo {volume} {19}} (\bibinfo {year} {2023})}\BibitemShut {NoStop}%
\bibitem [{\citenamefont {Kuz’min}\ \emph {et~al.}(2020)\citenamefont
  {Kuz’min}, \citenamefont {Skokov}, \citenamefont {Diop}, \citenamefont
  {Radulov},\ and\ \citenamefont {Gutfleisch}}]{Kuzmin2020}%
  \BibitemOpen
  \bibfield  {author} {\bibinfo {author} {\bibfnamefont {M.~D.}\ \bibnamefont
  {Kuz’min}}, \bibinfo {author} {\bibfnamefont {K.~P.}\ \bibnamefont
  {Skokov}}, \bibinfo {author} {\bibfnamefont {L.~V.~B.}\ \bibnamefont {Diop}},
  \bibinfo {author} {\bibfnamefont {I.~A.}\ \bibnamefont {Radulov}}, \ and\
  \bibinfo {author} {\bibfnamefont {O.}~\bibnamefont {Gutfleisch}},\ }\bibfield
   {title} {\enquote {\bibinfo {title} {Exchange stiffness of ferromagnets},}\
  }\href@noop {} {\bibfield  {journal} {\bibinfo  {journal} {The European
  Physical Journal Plus}\ }\textbf {\bibinfo {volume} {135}} (\bibinfo {year}
  {2020})}\BibitemShut {NoStop}%
\bibitem [{\citenamefont {Gurevich}\ and\ \citenamefont
  {Melkov}(1996)}]{Gurevich1996}%
  \BibitemOpen
  \bibfield  {author} {\bibinfo {author} {\bibfnamefont {A.}~\bibnamefont
  {Gurevich}}\ and\ \bibinfo {author} {\bibfnamefont {G.}~\bibnamefont
  {Melkov}},\ }\href {\doibase 10.1201/9780138748487} {\enquote {\bibinfo
  {title} {Magnetization oscillations and waves},}\ } (\bibinfo {year}
  {1996})\BibitemShut {NoStop}%
\bibitem [{\citenamefont {Sbiaa}\ \emph {et~al.}(2016)\citenamefont {Sbiaa},
  \citenamefont {Shaw}, \citenamefont {Nembach}, \citenamefont {Al~Bahri},
  \citenamefont {Ranjbar}, \citenamefont {Åkerman},\ and\ \citenamefont
  {Piramanayagam}}]{Sbiaa2016}%
  \BibitemOpen
  \bibfield  {author} {\bibinfo {author} {\bibfnamefont {R.}~\bibnamefont
  {Sbiaa}}, \bibinfo {author} {\bibfnamefont {J.~M.}\ \bibnamefont {Shaw}},
  \bibinfo {author} {\bibfnamefont {H.~T.}\ \bibnamefont {Nembach}}, \bibinfo
  {author} {\bibfnamefont {M.}~\bibnamefont {Al~Bahri}}, \bibinfo {author}
  {\bibfnamefont {M.}~\bibnamefont {Ranjbar}}, \bibinfo {author} {\bibfnamefont
  {J.}~\bibnamefont {Åkerman}}, \ and\ \bibinfo {author} {\bibfnamefont
  {S.~N.}\ \bibnamefont {Piramanayagam}},\ }\bibfield  {title} {\enquote
  {\bibinfo {title} {Ferromagnetic resonance measurements of {(Co/Ni/Co/Pt)}
  multilayers with perpendicular magnetic anisotropy},}\ }\href {\doibase
  10.1088/0022-3727/49/42/425002} {\bibfield  {journal} {\bibinfo  {journal}
  {Journal of Physics D: Applied Physics}\ }\textbf {\bibinfo {volume} {49}},\
  \bibinfo {pages} {425002} (\bibinfo {year} {2016})}\BibitemShut {NoStop}%
\bibitem [{\citenamefont {Nembach}\ \emph {et~al.}(2011)\citenamefont
  {Nembach}, \citenamefont {Silva}, \citenamefont {Shaw}, \citenamefont
  {Schneider}, \citenamefont {Carey}, \citenamefont {Maat},\ and\ \citenamefont
  {Childress}}]{Nembach2011}%
  \BibitemOpen
  \bibfield  {author} {\bibinfo {author} {\bibfnamefont {H.~T.}\ \bibnamefont
  {Nembach}}, \bibinfo {author} {\bibfnamefont {T.~J.}\ \bibnamefont {Silva}},
  \bibinfo {author} {\bibfnamefont {J.~M.}\ \bibnamefont {Shaw}}, \bibinfo
  {author} {\bibfnamefont {M.~L.}\ \bibnamefont {Schneider}}, \bibinfo {author}
  {\bibfnamefont {M.~J.}\ \bibnamefont {Carey}}, \bibinfo {author}
  {\bibfnamefont {S.}~\bibnamefont {Maat}}, \ and\ \bibinfo {author}
  {\bibfnamefont {J.~R.}\ \bibnamefont {Childress}},\ }\bibfield  {title}
  {\enquote {\bibinfo {title} {Perpendicular ferromagnetic resonance
  measurements of damping and {L}andé \(g\)-factor in sputtered
  {(Co${}_{2}$Mn)${}_{1\ensuremath{-}x}$Ge${}_{x}$} thin films},}\ }\href
  {\doibase 10.1103/PhysRevB.84.054424} {\bibfield  {journal} {\bibinfo
  {journal} {Phys. Rev. B}\ }\textbf {\bibinfo {volume} {84}},\ \bibinfo
  {pages} {054424} (\bibinfo {year} {2011})}\BibitemShut {NoStop}%
\bibitem [{\citenamefont {Ziegler}, \citenamefont {Biersack},\ and\
  \citenamefont {Littmark}(1985)}]{ziegler1985}%
  \BibitemOpen
  \bibfield  {author} {\bibinfo {author} {\bibfnamefont {J.}~\bibnamefont
  {Ziegler}}, \bibinfo {author} {\bibfnamefont {J.}~\bibnamefont {Biersack}}, \
  and\ \bibinfo {author} {\bibfnamefont {U.}~\bibnamefont {Littmark}},\ }\href
  {https://books.google.fi/books?id=xclwQgAACAAJ} {\emph {\bibinfo {title} {The
  Stopping and Range of Ions in Solids}}},\ Stopping and ranges of ions in
  matter\ (\bibinfo  {publisher} {Pergamon},\ \bibinfo {year}
  {1985})\BibitemShut {NoStop}%
\bibitem [{\citenamefont {Dumont}\ \emph {et~al.}(2007)\citenamefont {Dumont},
  \citenamefont {Keller}, \citenamefont {Popova}, \citenamefont {Schmool},
  \citenamefont {Tessier}, \citenamefont {Bhattacharya}, \citenamefont {Stahl},
  \citenamefont {Da~Silva},\ and\ \citenamefont {Guyot}}]{Dumont2007}%
  \BibitemOpen
  \bibfield  {author} {\bibinfo {author} {\bibfnamefont {Y.}~\bibnamefont
  {Dumont}}, \bibinfo {author} {\bibfnamefont {N.}~\bibnamefont {Keller}},
  \bibinfo {author} {\bibfnamefont {E.}~\bibnamefont {Popova}}, \bibinfo
  {author} {\bibfnamefont {D.~S.}\ \bibnamefont {Schmool}}, \bibinfo {author}
  {\bibfnamefont {M.}~\bibnamefont {Tessier}}, \bibinfo {author} {\bibfnamefont
  {S.}~\bibnamefont {Bhattacharya}}, \bibinfo {author} {\bibfnamefont
  {B.}~\bibnamefont {Stahl}}, \bibinfo {author} {\bibfnamefont {R.~M.~C.}\
  \bibnamefont {Da~Silva}}, \ and\ \bibinfo {author} {\bibfnamefont
  {M.}~\bibnamefont {Guyot}},\ }\bibfield  {title} {\enquote {\bibinfo {title}
  {Tuning magnetic properties with off-stoichiometry in oxide thin films: {A}n
  experiment with yttrium iron garnet as a model system},}\ }\href@noop {}
  {\bibfield  {journal} {\bibinfo  {journal} {Phys. Rev. B}\ }\textbf {\bibinfo
  {volume} {76}},\ \bibinfo {pages} {104413} (\bibinfo {year}
  {2007})}\BibitemShut {NoStop}%
\bibitem [{\citenamefont {Noun}\ \emph {et~al.}(2010)\citenamefont {Noun},
  \citenamefont {Popova}, \citenamefont {Bardelli}, \citenamefont {Dumont},
  \citenamefont {Bertacco}, \citenamefont {Tagliaferri}, \citenamefont
  {Tessier}, \citenamefont {Guyot}, \citenamefont {Berini},\ and\ \citenamefont
  {Keller}}]{Noun2010}%
  \BibitemOpen
  \bibfield  {author} {\bibinfo {author} {\bibfnamefont {W.}~\bibnamefont
  {Noun}}, \bibinfo {author} {\bibfnamefont {E.}~\bibnamefont {Popova}},
  \bibinfo {author} {\bibfnamefont {F.}~\bibnamefont {Bardelli}}, \bibinfo
  {author} {\bibfnamefont {Y.}~\bibnamefont {Dumont}}, \bibinfo {author}
  {\bibfnamefont {R.}~\bibnamefont {Bertacco}}, \bibinfo {author}
  {\bibfnamefont {A.}~\bibnamefont {Tagliaferri}}, \bibinfo {author}
  {\bibfnamefont {M.}~\bibnamefont {Tessier}}, \bibinfo {author} {\bibfnamefont
  {M.}~\bibnamefont {Guyot}}, \bibinfo {author} {\bibfnamefont
  {B.}~\bibnamefont {Berini}}, \ and\ \bibinfo {author} {\bibfnamefont
  {N.}~\bibnamefont {Keller}},\ }\bibfield  {title} {\enquote {\bibinfo {title}
  {Determination of yttrium iron garnet superexchange parameters as a function
  of oxygen and cation stoichiometry},}\ }\href@noop {} {\bibfield  {journal}
  {\bibinfo  {journal} {Phys. Rev. B}\ }\textbf {\bibinfo {volume} {81}},\
  \bibinfo {pages} {054411} (\bibinfo {year} {2010})}\BibitemShut {NoStop}%
\end{thebibliography}%

\end{document}